\documentclass[aps,amsmath,amssymb,reprint]{revtex4-2}

\usepackage{graphicx}
\usepackage{dcolumn}
\usepackage{bm}

\usepackage[utf8]{inputenc}
\usepackage[T1]{fontenc}
\usepackage{mathptmx}
\usepackage{etoolbox}

\usepackage{latexsym}
\usepackage{calligra}
\usepackage{mathtools}

\usepackage{hyperref}
\usepackage{color}

\definecolor{darkred}{rgb}{0.5,0,0}

\newcommand\DN[1]{\text{#1}}
\newcommand{\der}[2]{\dfrac{d #1}{d #2}}

\newcommand{\parder}[2]{\dfrac{\partial #1}{\partial #2}}
\newcommand{\pardern}[3]{\dfrac{\partial^{#3} #1}{\partial #2^{#3}}}

\newcommand{\grad}[1]{\nabla #1}

\begin{document}

\title[Leidenfrost drop dynamics]{Leidenfrost drop dynamics: \\ An approach to follow the complete evolution}

\author{Ren\'e Ledesma-Alonso}
 \email{rene.ledesma@ciencias.unam.mx}
 \affiliation{Facultad de Ciencias, Universidad Nacional Aut\'onoma de M\'exico, Av. Universidad 3000, Circuito Exterior S/N, Alcaldı\'ia Coyoac\'an, C.P. 04510 Ciudad Universitaria, Ciudad de M\'exico, M\'exico.}
\author{Benjamin Lalanne}
 \email{benjamin.lalanne@ensiacet.fr}
\affiliation{Laboratoire de G\'enie Chimique, Universit\'{e} de Toulouse, CNRS, INP, UPS, Toulouse, France.}%
\author{Jes\'us Israel Mor\'an-Cort\'es}
 \affiliation{Instituto de F\'isica, Benem\'erita Universidad Aut\'onoma de Puebla, Apartado Postal J-48, C.P. 72570, Puebla, M\'exico.}%
\author{Mart\'in Aguilar-Gonz\'alez}
 \affiliation{Instituto de F\'isica, Benem\'erita Universidad Aut\'onoma de Puebla, Apartado Postal J-48, C.P. 72570, Puebla, M\'exico.}%
\author{Felipe Pacheco-V\'azquez}
 \email{fpacheco@ifuap.buap.mx}
 \affiliation{Instituto de F\'isica, Benem\'erita Universidad Aut\'onoma de Puebla, Apartado Postal J-48, C.P. 72570, Puebla, M\'exico.}%

\date{\today}

\begin{abstract}
A new model to follow the complete evolution of a drop in Leidenfrost state is presented in this work.
The main ingredients of the phenomenon were considered, including: 1) the shape and weight of a sessile drop, according to its size, compared to the capillary length, using the Young-Laplace equation; 2) the evaporation at the entire surface of the drop, due to the heat transfer across the vapor film, to the proximitiy of a hot plate and to the diffusion in air; 3) the velocity, pressure and temperature fields at the vapor film, between the drop and the hot plate, which are recovered by means of a Hankel transform method, being valid for any size of drops and any thickness of vapor films (below the vapor film stability threshold); 4) an estimation of the thermo-capillary Marangoni convection flow, without simulating numerically the flow within the drop.
The aforementioned features were addressed and calculated, in order to include their effect within a single non-linear ODE, describing the temporal evolution of the size of the drop, through the Bond number.
Three dimensionless parameters, relating the thermophysical properties of the drop fluid and the surrounding air, control the development of the phenomenon.
All those properties were calculated according to the ideal gas approximation and to widely used empirical correlations, without any fitting parameter.
The model predictions were compared against experimental results, using different organic and inorganic compounds, for which a good agreement has been found, when no bounce or rotation of the drop spontaneously occurs
\end{abstract}

\maketitle

\section{Introduction}
\label{Sec:Intro}

The first observations of a liquid drop levitating over a hot plate, due to the generation of a vapor cushion beneath the drop, were made more than 250 years ago~\cite{Leidenfrost1756}.
Nowadays, a very large number of documents addressing different phenomena related to the Leidenfrost effect on drops can be found in the literature~\cite{Quere2013,Kabov2021,Stewart2022}.
Self-propulsion~\cite{Linke2006,Quere2011,Cousins2012,Colinet2017,Quere2018,Lohse2019,Matsumoto2021}, jumping~\cite{Pomeau2012,Chen2014,Sun2019,Graeber2021}, shape oscillation~\cite{Caswell2014,Pomeau2014,Burton2017,Burton2018,Quere2021}, and multi-component interaction~\cite{Raufaste2016,Pacheco2021,Sun2021} are a few examples of the topics that had been derived from the study of single drops in the Leidenfrost state.

In the past two decades, several attempts had been performed to predict the complete dynamics of Leidenfrost drops, considering direct numerical simulations~\cite{Xu2013,Rueda2016,Rueda2017,Guo2019,Chakraborty2022,Mialhe2023} and analytic models~\cite{Biance2003,Myers2009,Shi2019,Cai2020}.
Numerical simulations solve for the motion and temperature distribution both within the vapor and the internal liquid, by means of multiphase techniques, phase change models and by possibly taking into account more complex effects such as Marangoni convection.
They are particularly challenging to perform, and imply significant computation power and time, due to the multiscale nature of the problem, since the vapor film between the drop and the hot plate can be several orders of magnitude smaller than the drop. Therefore, extremely fine spatial grids are required in the film region, as well as the use of very small time steps to capture the drop dynamics, for stability purpose of the numerical method.
Thus, functionality of such numerical simulations depends as much on the mesh density, within the vapor domain and at the liquid-vapor interface, as on the reliability of the numerical methods to capture a two-phase flow with jump conditions at the interface, related to phase change.
However, accurate numerical simulations are highly valuable to evaluate the relative importance of different phenomena in the dynamics of the levitating droplets, through comparisons with experimental data.
For instance, recent numerical simulations from ~\citet{Mialhe2023} have revealed that thermo-capillary effects are important for setting the film thickness between the drop and the substrate: the Marangoni convection, induced by temperature gradients along the interface, induces a significant internal circulation in the droplet, which favors the drainage rate of the film, making it to be thinner when this effect is accounted for, consistently with experimental results~\cite{Pomeau2012}. 

Such comparisons, to validate the impact of different phenomena, can also be performed between experiments and simplified theoretical models that describe the shape of the drop and the motion of the vapor, which are also highly interesting due to the wide range of thermophysical mechanisms that are involved in the phenomenon.
In this way, some theoretical studies focus on the geometry, as well as the stability of the vapor film~\cite{Snoeijer2009,Sobac2014,Maquet2015}, considering a snapshot of the drop at a given time during its evaporation.
A precise instantaneous description of the liquid-vapor system is obtained by simultaneously solving the Young-Laplace equation and the lubrication approximation, while considering a constant evaporation flux. For instance, good agreement is observed on the geometrical characteristics of the vapor layer between prediction, from the theoretical approach of ~\citet{Sobac2014}, and experimental measurements, by interferometry of the bottom of water drops by ~\citet{Burton2012}.
Additionally, a main result of these kind of works is the maximum size, a radial extent of around 3.95 times the capillary length, before the vapor layer becomes unstable and vapor chimneys disturb the shape of the drop~\cite{Snoeijer2009}.

Other studies are directed towards the dynamics of Leidenfrost drops, considering the different heat transfer mechanisms that are involved.
Based on a quasi-static approach, time-dependent scaling laws for the radius of the drops and the thickness of the vapor films were developed for the first time by~\citet{Biance2003}.
The drop radius $r_{\text{max}}$ and the vapor film thickness $h$ evolve with time $t$, respectively as:
\begin{subequations}
\begin{align}
r_{\text{max}}(t)&=r_{\text{max},0}\left(1-\dfrac{t}{\mathcal{T}}\right)^n \ , \\
h(t)&=h_0\left(1-\dfrac{t}{\mathcal{T}}\right)^m \ ,
\label{eq:powlaw}
\end{align}
\end{subequations}
where $\mathcal{T}$ is the lifetime of the drop and $h_0$ is the initial film thickness, both related to the initial drop radius $r_{\text{max},0}$, the hot plate temperature $T_{\text{p}}$ and the thermophysical properties of the fluid.
For puddles, which are drops with an extent above the capillary length, the exponents take the values $n=2$ and $m=n/2$, whereas for marbles, which are quasi-spherical drops with a radius below the capillary length, $n=1/2$ and $m=4n/3$.
It is also important to denote that $\mathcal{T}$ and $h_0$ have different definitions, according to the size of the drop, for instance $\mathcal{T}\sim r_{\text{max},0}^{1/n}$.
In the aforementioned study, the evaporation and phase change at the base of the drops was considered for all the drop sizes, whereas the evaporation from the entire drop surface was only considered for marble drops.
This characterization with decoupled regimes avoids following the complete dynamics of an above-millimeter-sized drop, from its deposition on the hot plate till its full evaporation, with a continuous transition between the dynamics of a puddle and a marble.

Two recent works address the phenomenon based on simple arguments and ideas, for large drops with extents above the capillary length, both pointing towards power laws with the same description given by eqs.~\eqref{eq:powlaw}.
The first is a theoretical analysis with strong simplifications on the drop shape~\cite{Ziese2019}, considering a hemispherical geometry, which leads to different values of the exponents, \emph{i.e.} $n=4/5$ and $m=n/4$, also with $\mathcal{T}\sim r_{\text{max},0}^{1/n}$ and additional fitting results that are in the range $n\in\left(0.9,1.5\right)$.
Unfortunately, this model leads to estimates of the vapor thermal conductivity that are at least 3 orders of magnitude smaller than the values found in the literature.
The second is an experimental analysis that, based on the observations and a linear fitting of the drop mass in terms of the area projected over the hot plate~\cite{Orzechowski2021}, leads to $n=1$ and $\mathcal{T}\sim r_{\text{max},0}^{1/n}$ as well.
Besides the dependence on several fitting parameters, this model yields an overall heat transfer coefficient that varies with the size of the drop and, as a consequence, with time.
It is worth noting that both studies take into account the evaporation at the bottom of the drop, and the consequent generation of the vapor cushion, but neglect the evaporation from the top of the drop towards the surrounding air.
However, it has been shown that evaporation takes place over the entire surface, and not only in the film, when the drop is small and quasi-spherical~\cite{Sobac2015}, even if the contribution from the vapor film is largely dominant in the case of puddles.
Both phenomena should therefore be taken into account to develop a dynamic model of the evaporation of a drop in the Leidenfrost state.

Another two works attempt to follow the complete drop dynamics, through models that consider a quasi-static evolution of the drops~\cite{Myers2009,Cai2020}.
The first employs the lubrication approximation and pure heat conduction at the vapor layer, which is considered to have a uniform thickness.
Relatively large drops are considered in this work, starting from initial radii of around 3.95 times the capillary length.
Two fitting parameters are employed, \emph{i.e.} a substrate-to-vapor heat transfer coefficient and an evaporation rate~\cite{Myers2009}.
The second also includes the lubrication approximation, and incorporates heat conduction and radiation at the vapor layer, which is considered to have a thickness that varies with the radial position, as a previous quasi-static study~\cite{Sobac2014}.
Additionally, mass diffusion is considered for the top region of the drops, but considering a spherical drop approach and the gas around the upper part of the drop as a pure vapor phase ~\cite{Cai2020}, which may not be the case in experiments.
Besides, despite the completeness of this work, it is limited by reduced extents of the drops, which requires initial radii either around or below the capillary length.

Evidently, in all the aforementioned studies~\cite{Ziese2019,Orzechowski2021,Myers2009,Cai2020}, there is partial agreement between the experimental results and the developed models, either because of the reduced range of application (small or large drops) or the employment of fitting parameters.
To date, different power laws had been presented, and several models and procedures that work on limited ranges had been developed, but still a common basis for the complete evolution of Leidenfrost drops, with relatively large initial radii (up to 3.5 times the capillary length), has not been established yet.

Among other outcomes of the phenomenon, the dependence of the lifetime of the drops on the different physical parameters, such as initial drop size and plate temperature, remains empirical.
In a recent study~\cite{Sobac2015}, considering pure conduction at the air surrounding the drop, the drop lifetime $\mathcal{T}$ has been overestimated with respect to experimental data.
Simultaneously, the authors propose an empiric power law for $\mathcal{T}$: 
\begin{align}
\mathcal{T}&=aV_0^n \ ,
\end{align}
where $V_0$ is the initial volume, while $a$ and $n$ are fitting parameters.
According to experimental results and a fitting procedure, $a$ is strongly dependent on the temperature of the plate $T_{\text{p}}$, whereas $n$ remains nearly constant.
For instance $a\approx8.9\times10^4$ s$\cdot$m$^{-3n}$ for $T_{\text{p}}=470\, ^{\circ}\text{C}$, while $a\approx1.6\times10^5$ s$\cdot$m$^{-3n}$ for $T_{\text{p}}=300\, ^{\circ}\text{C}$, whereas $n\approx 0.4$ throughout the same temperature range.
In another work from the same authors, the following power law is proposed to assess the dependence of the drop lifetime on the temperature difference~\cite{Maquet2015}:
\begin{align}
\mathcal{T}&=A\left(T_{\text{p}}-T_{\text{l}}\right)^{-3/4} \ ,
\end{align}
where $T_{\text{l}}$ is the drop temperature and $A\approx4.6\times10^4$ s$\cdot^{\circ}$C$^{3/4}$ is a fitting parameter for the same initial volume $V_0\approx1.5\times10^{-4}$ m$^3$.
This relation shows a very good agreement with experimental data when increased gravity conditions are implemented and the puddle regime is observed for small volume drops, but, once more, overestimates the experimental lifetime of drops at normal gravity conditions.

An estimation of the drop evolution and lifetime is essential for many applications, including friction reduction~\cite{Hashmi2012,Segawa2018}, microfluidics~\cite{Park2011,Adera2013,Dodd2016,Dodd2020}, and cooling techniques~\cite{Mudawar2017a,Mudawar2017b}, among other possible usages that are still unveiled.
Therefore, further research is required to provide a complete description of the dynamics of Leidenfrost drops.

In this study, we revisit the Leidenfrost effect, with the aim of developing a simple but effective strategy to estimate outputs, for instance the drop size dependence on time and its lifetime, in terms of the input parameters, such as the hot plate temperature and the thermophysical properties of the fluid.

Considering that the flow and heat transfer at the vapor film occurs at a much shorter time scale than the size reduction of the drop due to the evaporation, a quasi-static approximation is employed to develop a complete dynamic model.
For an axisymmetric configuration the following procedure is performed: 1) the Young-Laplace equation is solved to find the shape of the drop for a given volume; 2) a Stokes flow is obtained for the vapor film, that is generated from the evaporation at the bottom of the drop and is expelled radially due to the weight of the drop; 3) the temperature distribution at the vapor film is obtained from heat conduction; 4) the evaporation rate is deduced from the Stefan condition at the bottom of the drop, from the heat transfer at the lateral region of the lower hemisphere, and from the Stefan convective evaporation flow at the top hemisphere of the drop; 5) a mass balance is applied to follow the evolution of the drop.
Compared to previous modelling approaches of the phenomena ~\cite{Myers2009, Cai2020}, there is no lubrication assumption for the flow in the vapor film, the analytical model being thus valid without any limiting value of the film thickness compared to the drop radius, which allows to deal continuously with both puddles and quasi-spherical drops.
Besides, different contributions for evaporation are systematically considered in the model (within the vapor film, in the upper part of the drop and at its lateral region), without assuming a pure vapor for the ambient gas in the neighboring of the northern drop hemisphere.
This approach allows us to study the dynamics of a drop in Leidenfrost state, from its initial size and shape (above or below the capillary threshold) until it disappears due to its full evaporation.
Experiments were carried out using seven fluids, to serve as a basis for assessing the relevance of the analytical model that is proposed.
The latter is written in dimensionless form and can be easily extended, if further contributions need to be included such as radiation or natural convection effects, which have not been taken into account in the range of temperature explored in this investigation.
It is important to mention that no fitting parameters are employed in the proposed methodology.

\section{Theoretical analysis}
\label{Sec:Theo}

Consider a drop of volume $V$, levitating over a vapor film of thickness $h$, within a cylindrical coordinate system with axisymmetric geometry.
$T_{\text{p}}$ and $T_{\text{l}}$ are the temperatures of the hot plate and that of the liquid, respectively.
Due to the relatively slow evaporation process~\cite{Biance2003,Sobac2015}, compared to the vapor film dynamics which is also slow, a quasi-static approximation for the shape of the drop, the vapor flow below the drop and the temperature distribution can be regarded.
The effect of time will be considered in the dynamics of the drop volume $V(t)$ and the vapor film thickness $h(t)$.
Drops in the Leidenfrost state are considered to be at the saturated liquid conditions, thus for any plate temperature $T_{\text{p}}$, the temperature of the drops $T_{\text{l}}$ is considere to be saturated liquid temperature.

\begin{figure*}[t]
\centering
\includegraphics[width=0.75\textwidth]{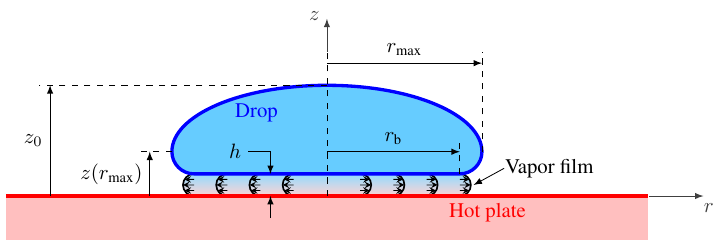}
\caption{
Leidenfrost drop schema.
The dimensions of the drop and vapor layer are shown, which introduction and description are given in section \ref{Sec:Shape}.
}
\label{fig:schema}
\end{figure*}

\subsection{A sessile drop with non-wetting condition, contact angle $\theta=0$}
\label{Sec:Shape}

Given a volume of a drop $V$, the equivalent radius $R$ of a sphere with the same volume is given by $V=(4\pi/3)R^3$.
Taking $R$ as the characteristic length of the drop, we recall the definitions of the capilary length $\lambda_{\text{c}}$ and Bond number $\DN{Bo}$:
\begin{subequations}
\begin{align}
\lambda_{\text{c}}&=\sqrt{\dfrac{\sigma}{\rho_{\text{l}} g}} \ , \\
\DN{Bo}&=\left(\dfrac{R}{\lambda_{\text{c}}}\right)^2 \ .
\label{DN:Bo}
\end{align}
\end{subequations}
where $\sigma$ is the surface tension of the fluid, $\rho_{\text{l}}$ is the density of the fluid in the liquid phase, and $g$ is the gravitational acceleration.

At equilibrium, the surface energy of the drop should be at a minimum, with the constraint of a fixed volume.
Thus, the shape of the drop is described by the Young-Laplace equation~\cite{deGennes}, which relates the vertical $z$ and radial $r(z)$ coordinates at the surface of the drop, for a given set of parameters, including $\sigma$ and $\rho_{\text{l}}$, as well as $\Delta P_0$ being the overpressure inside the drop at its top, which vertical position is $z_0$.
Additionally, one may define that the bottom of the drop is placed at the vertical position $h$, as depicted in Fig.~\ref{fig:schema}.
Now, defining $\xi=r/\lambda_{\text{c}}$ and $\eta=\left(z_0-z\right)/\lambda_{\text{c}}$, we can write the ODE that describes the drop shape in dimensionless terms:
\begin{align}
\dfrac{1}{\xi\left[1+\left(\dot{\xi}\right)^2\right]^{1/2}}
-\dfrac{\ddot{\xi}}{\left[1+\left(\dot{\xi}\right)^2\right]^{3/2}}
&=\kappa_0+\eta \ ,
\label{eq:ODE}
\end{align}
with $\kappa_0=(\lambda_{\text{c}}\, \Delta P_0)/\sigma$, whereas $\dot{\xi}$ and $\ddot{\xi}$ are the first and second derivatives of $\xi$ with respect to $\eta$.

\begin{figure*}[t]
\centering
\includegraphics[width=0.6\textwidth]{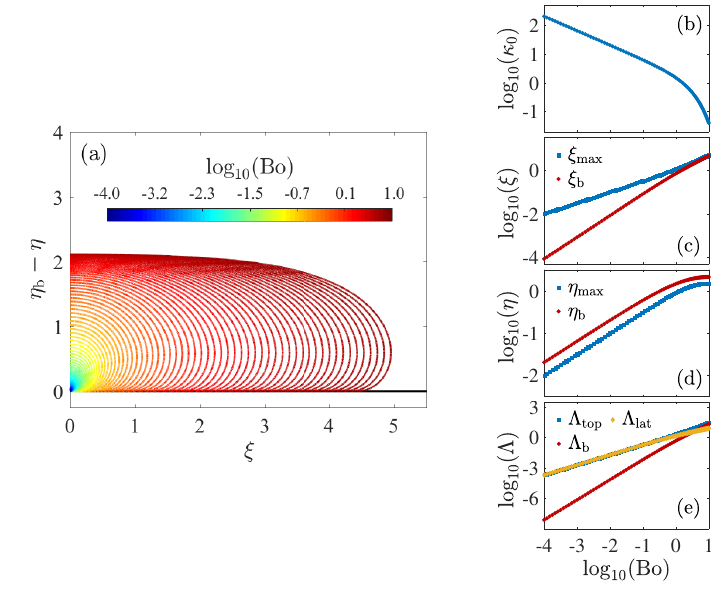} \\
\caption{
(a) Drop dimensionless shape for different values of the Bond number $\DN{Bo}$.
Geometric dimensionless properties of the drop as functions of $\DN{Bo}$:
(b) Curvature $\kappa_0$,
(c) Radial positions $\xi$,
(d) Axial positions $\eta$, and
(e) Surface areas $\Lambda$.
}
\label{fig:shape}
\end{figure*}

The shape of the drop $\xi(\eta)$ is determined by solving numerically eq.~\eqref{eq:ODE}, for a given value of $\kappa_0$.
Afterwards, the geometric properties, such as the maximum radius $r_{\text{max}}=\lambda_{\text{c}}\xi_{\text{max}}$, the bottom radius $r_{\text{b}}=\lambda_{\text{c}}\xi_{\text{b}}$, the drop height $\left(z_0-h\right)=\lambda_{\text{c}}\eta_{\text{b}}$, and the vertical distance from the top to the position of maximum radius $\left[z_0-z\left(r_{\text{max}}\right)\right]=\lambda_{\text{c}}\eta_{\text{max}}$ can be determined.
As well, the volume of the droplet and the Bond number are computed as follows:
\begin{subequations}
\begin{align}
V&
=\pi\, \lambda_{\text{c}}^3\int_{0}^{\eta_{\text{b}}} \left[\xi(\eta)\right]^2\, d\eta \ , \\
\DN{Bo}&=\dfrac{1}{\lambda_{\text{c}}^2}\left(\dfrac{3V}{4\pi}\right)^{2/3} \ .
\end{align}
\end{subequations}
The surface of the drop that is in contact with the supporting vapor film reads:
\begin{subequations}
\begin{align}
A_{\text{b}}&
=\pi\, \lambda_{\text{c}}^2 \xi_{\text{b}}^2 \ , \\
\Lambda_{\text{b}}&=\dfrac{A_{\text{b}}}{\pi\, \lambda_{\text{c}}^2} \ .
\end{align}
\end{subequations}
The lateral region of the lower hemisphere, from $h$ to $z(r_{\text{max}})$, has an area that is computed as follows:
\begin{subequations}
\begin{align}
A_{\text{lat}}&
=2\pi\, \lambda_{\text{c}}^2\int_{\eta_{\text{max}}}^{\eta_{\text{b}}} \xi(\eta)\sqrt{1+\left(\dfrac{d\xi}{d\eta}\right)^2}\, d\eta \ ,
\\
\Lambda_{\text{lat}}&=\dfrac{A_{\text{lat}}}{\pi\, \lambda_{\text{c}}^2} \ ,
\end{align}
\end{subequations}
whereas the surface that is in contact with the air above the drop, from $z(r_{\text{max}})$ to $z_0$, is given by:
\begin{subequations}
\begin{align}
A_{\text{top}}&
=2\pi\, \lambda_{\text{c}}^2\int_0^{\eta_{\text{max}}} \xi(\eta)\sqrt{1+\left(\dfrac{d\xi}{d\eta}\right)^2}\, d\eta \ ,
\\
\Lambda_{\text{top}}&=\dfrac{A_{\text{top}}}{\pi\, \lambda_{\text{c}}^2} \ .
\end{align}
\end{subequations}
The shape of the drop and the described geometric properties are presented in Fig.~\ref{fig:shape}, for different values of the Bond number in the range $\DN{Bo}\in\left[10^{-4},10^1\right]$.

\subsection{Vapor motion and temperature}

The vapor film that is present below the drop is fed by the liquid-vapor phase change and expelled by the weight of the drop.
Since the thickness of the film $h$ is expected to be small, the effects of the gravity can be neglected, an axisymmetric, steady-state and viscous flow (small Reynolds number) and a conduction dominated heat transfer (small Peclet number) can be considered within the film~\cite{Sobac2015}.
Under this circumstance, the mass, momentum and energy differential balances describe the behavior the radial $\upsilon_r$ and axial $\upsilon_z$ velocity components, the pressure $P$ and the temperature $T$ within the film, according to the values of the dynamic viscosity $\mu_{\text{v}}$ and thermal diffusivity $\alpha_{\text{v}}$ of the vapor, at the film temperature $T_{\text{v}}=(T_{\text{p}}+T_{\text{l}})/2$.

Now, let us define the dimensionless variables:
\begin{subequations}
\begin{align}
y&=\dfrac{z}{h} \ , &
x&=\dfrac{r}{r_{\text{b}}} \ , &
w&=\dfrac{\upsilon_z}{\upsilon_0} \ , \\
u&=\dfrac{h\, \upsilon_r}{r_{\text{b}}\, \upsilon_0} \ , &
p&=\dfrac{h^3\, P}{\mu_{\text{v}}\, \upsilon_0\, r_{\text{b}}^2} \ , & 
\theta&=\dfrac{T_{\text{p}}-T}{T_{\text{p}}-T_{\text{l}}} \ ,
\end{align}
as well as the aspect ratio $H$, which reads:
\begin{align}
H&=\dfrac{h}{r_{\text{b}}} \ .
\end{align}
\end{subequations}
where $\upsilon_0$ is the characteristic vertical speed of the vapor, expelled from the bottom of the drop at $r=0$.
Using them, we transform the mass, momentum (radial and axial) and energy balances into the following dimensionless system of PDEs:
\begin{subequations}
\begin{align}
0&=\dfrac{1}{x}\parder{\left(x u\right)}{x}+\parder{w}{y} \ , \label{eq:m} \\
\parder{p}{x}&=H^2\left(\pardern{u}{x}{2}+\dfrac{1}{x}\parder{u}{x}-\dfrac{u}{x^2}\right)+\pardern{u}{y}{2} \ , \label{eq:r} \\
\dfrac{1}{H^2}\parder{p}{y}&=H^2\left(\pardern{w}{x}{2}+\dfrac{1}{x}\parder{w}{x}\right)+\pardern{w}{y}{2} \ , \label{eq:a} \\
0&=H^2\left(\pardern{\theta}{x}{2}+\dfrac{1}{x}\parder{\theta}{x}\right)+\pardern{\theta}{y}{2} \ , \label{eq:e}
\end{align}
\label{eq:mrae}
\end{subequations}
respectively.

Eqs.~\eqref{eq:mrae} must be solved, subjected to the boundary conditions:
\begin{subequations}
\begin{align}
\text{at} \quad y&=0 \quad
\rightarrow
\begin{cases}
u&=0 \ , \\
w&=0 \ , \\ 
\theta&=0 \\ 
\end{cases}
\\[1ex]
\text{at} \quad y&=1 \quad
\rightarrow
\begin{cases}
u&=f_{\text{M}}(x) \ , \\
w&=-f_{\upsilon}(x) \ ,\\ 
\theta&=f_{\theta}(x) 
\end{cases}
\end{align}
\label{eq:BCs}
\end{subequations}
where $f_{\text{M}}(x)$, $f_{\upsilon}(x)$ and $f_{\theta}(x)$ are dimensionless functions that describe the interface radial velocity (due to Marangoni convection within the drop), the radial distribution of vapor generation and the temperature variation, respectively, at the top boundary $y=1$.
These functions are described by: 
\begin{subequations}
\begin{align}
f_{\text{M}}(x)&=u_{\text{M}}\,  x^2\exp\left(-4\pi x^2\right) \ ,
\\[1ex]
f_{\upsilon}(x)&=\mathcal{H}\left(1-x\right) \ ,
\\[1ex]
f_{\theta}(x)&=\begin{cases}
1 & \text{for } 0\leq x\leq 1 \ , \\
1/x &\text{for } 1<x \ ,
\end{cases}
\end{align}
\label{eqs:dimfun}
\end{subequations}
where $\mathcal{H}(\cdot)$ is the Heaviside step function.
In turn, $u_{\text{M}}$ is the normalized intensity of the Marangoni convection flow, which is given by:
\begin{align}
u_{\text{M}}&=\dfrac{H\, U_{\text{M}}}{\upsilon_0}\ , &
U_{\text{M}}&=\dfrac{\alpha_{\text{v}}}{ \lambda_{\text{c}}}\, \DN{Ma}_{\text{v}} \ ,
\label{eq:mgnspd}
\end{align}
where $\DN{Ma}$ is the thermal Marangoni number, defined as:
\begin{align}
\DN{Ma}_{\text{v}}=-\left(\der{\sigma}{T}\bigg\vert_{T_{\text{l}}}\right)\dfrac{\left(T_{\text{l}}-T_{\text{int}}\right)}{\mu_{\text{v}}}\dfrac{\lambda_{\text{c}}}{\alpha_{\text{v}}} \ .
\end{align} 
Further details on the dimensionless functions, introduced in eqs.~\eqref{eqs:dimfun}, are presented in the Supplementary Material (SM)~\cite{SuppMat}.

\begin{figure*}[t]
\centering
\includegraphics[width=0.75\textwidth]{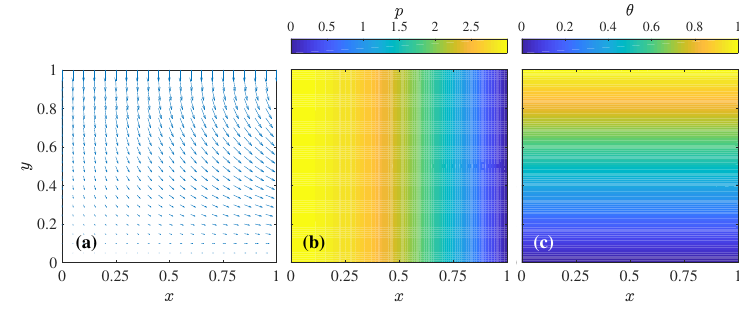} \\
\caption{
Dimensionless fields: (a) velocity, (b) pressure, and (c) temperature; resulting from the solution of the mass, momentum and energy balances, described by eqs.~\eqref{eq:mrae}, for $\DN{Bo}_0=7.185$, $H=0.017$ and $\DN{Ma}_{\text{v}}/\DN{Ja}_{\text{v}}=0$.
}
\label{fig:Fields}
\end{figure*}

Applying the Hankel transform~\cite{Sneddon} of order 0 to eqs.~\eqref{eq:m}, \eqref{eq:a}, and \eqref{eq:e}, whereas the 1st order Hankel transform is applied to the eq.~\eqref{eq:r}, yields a system of coupled ODEs in terms of the Hankel transform equivalents of the velocity components, pressure and temperature.
The system of ODEs can be solved analytically, thus the aforementioned fields are recovered, once the corresponding inverse Hankel transform has been performed.
Further details on the complete Hankel transfrom procedure are presented in SM~\cite{SuppMat}.
An example of the dimensionless fields, velocity, pressure and temperature, is presented in Fig.~\ref{fig:Fields}.

Note that the lubrication approximation had not been considered in the present study, in order to ensure a smooth transition from the large drop towards the small drop regime.
In previous works~\cite{Pomeau2012,Sobac2014,Kabov2017}, it has been observed that the lubrication regime is no longer valid for small spherical drops, with $\xi_{\text{max}}<1$, which size becomes comparable to the film thickness, for instance when $H\sim 0.1$.

\subsection{Evaporation}

Evaporation takes place at the surface of the drop, but driven by different mechanisms.
It occurs:
\begin{enumerate}
\item below the drop, at $r\leq r_{\text{b}}$ with $z=h$, due to the heat transfer through the thin film that supports the drop, feeding it with vapor.
\item at the lateral region of the lower hemisphere, between the film region and the radial maximum extent, at $r_{\text{b}}<r\leq r_{\text{max}}$ with $z\leq z\left(r_{\text{max}}\right)$, due to the proximity of the hot plate and the corresponding heat transfer.
\item above the drop, at $0<r\leq r_{\text{max}}$ with $z\left(r_{\text{max}}\right)<z$, due to the diffusion process of liquid molecules in the surrounding air, at relatively high temperatures.
\end{enumerate}
In this section, the three mechanisms are considered and analized.

\subsubsection{Vapor film below the drop}

Below the drop, at its interface with the vapor film, the evaporation flux $\Phi_{\text{b}}$ is defined from the Stefan condition, which can be written as:
\begin{align}
\Phi_{\text{b}}&=-\dfrac{2 k_{\text{v}}}{\rho_{\text{l}} L_{\text{lv}} r_{\text{b}}^2}\int_0^{r_{\text{b}}} \left(\parder{T}{z}\bigg\vert_{z=h}\right)\, r\, dr \ ,
\end{align}
where $L_{\text{lv}}$ is the latent heat of the liquid-vapor phase change, whereas $k_{\text{v}}$ is the thermal conductivity of the vapor, at the film temperature $T_{\text{v}}$.
In dimensionless terms, this expression reads:
\begin{subequations}
\begin{align}
\dfrac{h\, \Phi_{\text{b}}}{\alpha_{\text{v}}}&=2\, \left(\dfrac{\rho_{\text{v}}}{\rho_{\text{l}}}\right)\, \DN{Ja}_{\text{v}}\, G_{\theta} \ , \\
G_{\theta}(H)&=\int_0^1 \left(\parder{\theta}{y}\bigg\vert_{y=1}\right)\, x\, dx \ ,
\end{align}
\label{eq:Gth}
\end{subequations}
where we recall the Jakob number $\DN{Ja}_{\text{v}}$ and the thermal diffusivity $\alpha_{\text{v}}$ , defined as follows:
\begin{subequations}
\begin{align}
\DN{Ja}_{\text{v}}&=\dfrac{c_{p\text{v}}\left(T_{\text{p}}-T_{\text{l}}\right)}{L_{\text{lv}}} \ , \\
\alpha_{\text{v}}&=\dfrac{k_{\text{v}}}{\rho_{\text{v}}\, c_{p\text{v}}} \ ,
\end{align}
\end{subequations}
with $\rho_{\text{v}}$ and $c_{p\text{v}}$ being the density and the specific heat capacity of the vapor at constant pressure, rexpectively, both at the temperature $T_{\text{v}}$.

For a quasi-static situation, the velocity of the expelled vapor $\upsilon_0$ is related to the evaporation flux $\Phi_{\text{b}}$ by means of the mass conservation relation: 
\begin{align}
\upsilon_0&=\left(\dfrac{\rho_{\text{l}}}{\rho_{\text{v}}}\right)\, \Phi_{\text{b}} \ .
\label{eq:masscons}
\end{align}


\subsubsection{Lateral region of the lower hemisphere}

At the lower hemisphere of the drop, at the lateral region outside from the vapor film region, but before reaching $r_{\text{max}}$, there is evaporation due to the proximity of the hot plate.
It has been verified in previous studies, that an important contribution to the total evaporation comes from this region~\cite{Sobac2015}.
This evaporation flux $\Phi_{\text{lat}}$ is computed as follows:
\begin{align}
\Phi_{\text{lat}}&=-\dfrac{2\pi k_{\text{v}}}{\rho_{\text{l}} L_{\text{lv}} A_{\text{lat}}}\int_{r_{\text{b}}}^{r_{\text{max}}} \left(\grad{T}\cdot\widehat{n}\right)\, r\, dr \ ,
\end{align}
where $\grad{T}\cdot\widehat{n}$ is the directional derivative in the direction of the unit normal vector $\widehat{n}$.
The integral is estimated by a temperature profile that is inversely porportional to the distance (spherical symmetry) in the perpendicular direction to the drop surface, being $T=T_{\text{l}}$ at the surface of the drop and $T=T_{\text{p}}$ at the plate:
\begin{subequations}
\begin{align}
\int_{r_{\text{b}}}^{r_{\text{max}}} \left(\grad{T}\cdot\widehat{n}\right)\, r\, dr
&\approx\int_{r_{\text{b}}}^{r_{\text{max}}} \left(\dfrac{T_{\text{l}}-T_{\text{p}}}{r}\right) F_r \, r\, dr \ , \\[2ex]
F_r\left(r,z,\dot{z}\right)&=\dfrac{r}{z\left[1+\left(\dot{z}\right)^2\right]^{1/2}}+1 \ ,
\end{align}
\end{subequations}
where $z(r)$ describes the vertical position of the drop surface, at this region $r_{\text{b}}<r\leq r_{\text{max}}$, whereas $\dot{z}$ is the derivative with respect to $r$.
In dimensionless terms, this expression reads:
\begin{subequations}
\begin{align}
\dfrac{\lambda_c\, \Phi_{\text{lat}}}{\alpha_{\text{v}}}&=2\left(\dfrac{\rho_{\text{v}}}{\rho_{\text{l}}}\right)\dfrac{\DN{Ja}_{\text{v}}\, G_{\eta}}{\Lambda_{\text{lat}}} \ , \\[1ex]
G_{\eta}(\DN{Bo},H)&=-\int_{\xi_{\text{max}}}^{\xi_{\text{b}}}
f_{\xi} \, d\xi \ , \\[1ex]
f_{\xi}(\xi,\eta,\dot{\eta})&=
\dfrac{\xi}{\left[H \xi_{\text{b}}+\eta_{\text{b}}-\eta\right]\left[1+\left(\dot{\eta}\right)^2\right]^{1/2}}+1 \ .
\end{align}
\label{eq:Geta}
\end{subequations}
where $\eta\left(\xi\right)$ is the dimensionless vertical position and $\dot{\eta}$ is the derivative with respect to $\xi$.
Once more, the thermal diffusivity $\alpha_{\text{v}}$ is taken at the film temperature $T_{\text{v}}$.

\subsubsection{Diffusion in air above the drop}

Evaporation from the drop toward the surrounding air occurs at the top hemisphere of the drop, along the area $A_{\text{top}}$.
This evaporation flux $\Phi_{\text{top}}$ is computed as follows:
\begin{align}
\Phi_{\text{top}}&=\left(\dfrac{\rho_{\text{air}}}{\rho_{\text{l}}}\right)\, K_{lv}\, \left(\dfrac{\rho_{\text{int}}-\rho_{\infty}}{\rho_{\text{air}}-\rho_{\text{int}}}\right)
\notag \\
&\approx \left(\dfrac{\rho_{\text{air}}}{\rho_{\text{l}}}\right)\, K_{lv}\, \left(\dfrac{\rho_{\text{int}}}{\rho_{\text{air}}-\rho_{\text{int}}}\right) \ .
\label{eq:diff}
\end{align}
where $\rho_{\text{int}}$ and $\rho_{\infty}$ are the vapor densities at the air-liquid interface and far from the drop, respectively, $\rho_{\text{air}}$ is the air density at the reference temperature $T_{\text{ref}}=T_{\text{l}}+(T_{\text{p}}-T_{\text{l}})/3$, and $K_{lv}$ is a mass transfer coefficient.
In the case of an oblate spheroid within an infinite media, with major axis $2r_{\text{max}}$ and minor axis $2\left[z_0-z\left(r_{\text{max}}\right)\right]$, the mass transfer coefficient $K_{lv}$ can be estimated from the average Sherwood number, defined as:
\begin{align}
\DN{Sh}&=\dfrac{2r_{\text{max}}\, K_{lv}}{\mathcal{D}_{\text{v,air}}} \ ,
\end{align}
where $\mathcal{D}_{\text{v,air}}$ is the diffusion coefficient of the fluid into air, at the temperature of the drop $T_{\text{l}}$.

The Sherwood number considered herein is a modification from the spheroid transport process~\cite{Alassar1999}, that takes into account the Stefan convective flow, reading:
\begin{align}
\DN{Sh}&=\DN{Sh}_0 \, \dfrac{\ln\left(1+\DN{Bm}\right)}{\DN{Bm}} \ ,
\end{align}
with:
\begin{widetext}
\begin{equation}
\DN{Sh}_0\left(\omega\right)=\dfrac{-4}{\cosh\left(\omega\right) \left\{2 \tan^{-1}\left[\exp\left(\omega\right)\right]-\pi\right\} \left\{1+\sinh\left(\omega\right) \tanh\left(\omega\right) \ln\left[\coth\left(\omega/2\right)\right]\right\}} \ ,
\label{eq:Sh0}
\end{equation}
\end{widetext}
where $\omega=\tanh^{-1} \left(\eta_{\text{max}}/\xi_{\text{max}}\right)$, and a dependence on the Spalding mass transfer number $\DN{Bm}$ has been included.
This dimensionless number is given by:
\begin{align}
\DN{Bm}\approx\dfrac{\gamma_{\text{int}}}{1-\gamma_{\text{int}}} \ ,
\end{align}
with $\gamma_{\text{int}}=\rho_{\text{int}}/\rho_{\text{air}}$ being the mass fraction of vapor at the air-liquid interface, and assuming that the vapor density far from the drop is much smaller than that at the drop surface, \emph{i.e.} $\rho_{\text{int}}\gg \rho_{\infty}$.
Details on the determination of $\gamma_{\text{int}}$ are presented in SM~\cite{SuppMat}.

In dimensionless terms, the evaporation flux to the air is thus given by:
\begin{align}
\dfrac{\lambda_c \, \Phi_{\text{top}}}{\alpha_{\text{v}}}&=\left(\dfrac{\rho_{\text{v}}}{\rho_{\text{l}}}\right)\dfrac{\Pi_1^{\dagger}\, \DN{Sh}_0}{2\xi_{\text{max}}} \ ,
\label{eq:diff}
\end{align}
where we introduced the dimensionless group $\Pi_1^{\dagger}$, which reads:
\begin{align}
\Pi_1^{\dagger}&=\left(\dfrac{\rho_{\text{air}}\, \mathcal{D}_{\text{v,air}}}{\rho_{\text{v}}\, \alpha_{\text{v}}}\right)\, \ln\left(1+\DN{Bm}\right) \ .
\label{eq:Pi1}
\end{align}

Based on the computations of previous studies~\cite{Sobac2015}, the loss of mass of the droplet due to evaporation in the upper part can be significant for droplets of small size, for which the shape is almost spherical.

\subsection{Drop weight and vapor pressure}

In quasi-static conditions, the weight of the drop should be balanced by the integration of the pressure that is distributed at the vapor film-liquid interface.
This statement yields the equation:
\begin{align}
\rho_{\text{l}}\, g \, V&=2\pi \int_0^{r_{\text{b}}} \left(\Delta P\big\vert_{z=h}\right)\, r\, dr \ ,
\end{align} 
which, using the dimensionless variables and numbers, eq.~\eqref{eq:masscons} and the definition given in eq.~\eqref{eq:Gth}, becomes:
\begin{subequations}
\begin{align}
\dfrac{\Pi_2}{3}\, \DN{Bo}^{3/2}&=\dfrac{G_{\theta}\, G_{p}}{H^4} \ , \\[1ex]
G_{p}\left(H,u_{\text{M}}\right)&=\int_0^1 \left(p\big\vert_{y=1}\right)\, x\, dx \ ,
\end{align} 
\label{eq:HBo}
\end{subequations}
where we introduced the dimensionless group $\Pi_2$, reading:
\begin{align}
\Pi_2&=\left(\dfrac{\rho_{\text{l}}\, g\, \lambda_{\text{c}}^3}{\alpha_{\text{v}}\, \mu_{\text{v}}}\right)\, \dfrac{1}{\DN{Ja}_{\text{v}}} \ .
\label{eq:Pi2}
\end{align}
The dimensionless group $\Pi_2$ compares the characteristic weight of the drop $\rho_{\text{l}} g \lambda_c^3$ with the levitation capacity of the vapor, provided by the transport properties of the vapor $\alpha_{\text{v}}$ and $\mu_{\text{v}}$, as well as the liquid-vapor phase change by means of the Jakob number $\DN{Ja}_{\text{v}}$.

Eq.~\eqref{eq:HBo} must be solved to determine the thickness of the vapor film $H\left(\Pi_2,\DN{Bo},u_{\text{M}}\right)$, corresponding to a given set of values for the dimensionless group $\Pi_2$, the Bond number $\DN{Bo}$, and the dimensionless intensity of the Marangoni convection flow $u_{\text{M}}$.


\subsection{Marangoni convection flow}

For the approach presented in this work, an estimate of the Marangoni convection flow within the drop is taken into account, without resorting to a direct numerical simulation for the calculation of the velocity field within the drop.
Just the intensity of the Marangoni convection flow is required, as given by eq.~\eqref{eq:mgnspd}, which acts as a boundary condition for the flow at the vapor film.
Using eqs:~\eqref{eq:Gth} and \eqref{eq:masscons}, and the definition of the thermal Marangoni number, it can be written as follows:
\begin{align}
u_{\text{M}}&=\dfrac{H^2\, \xi_{\text{b}}\, \DN{Ma}_{\text{v}}}{2 G_{\theta}\, \DN{Ja}_{\text{v}}} \ .
\label{eq:uM}
\end{align}
Eq.~\eqref{eq:uM} is an additional relation that has to be considered in order to solve eq.~\eqref{eq:HBo}, since the value of $u_{\text{M}}$ influences the behavior of the pressure and velocity fields, $p$ and $(u,w)$, respectively.
As a consequence, $u_{\text{M}}$ has an effect over the dimensionless quantity $G_{p}$, and on the solution of eq.~\eqref{eq:HBo}, leading to a dependence of the thickness of the vapor film $H\left(\Pi_2,\DN{Bo},\DN{Ma}_{\text{v}}/\DN{Ja}_{\text{v}}\right)$.

\subsection{Drop size evolution over time}

The volume of the drop changes due to the evaporation processes that takes place at the different regions of the surface of the drop.
It follows a dynamics that is described by:
\begin{align}
\der{V}{t}&=-A_{\text{b}}\Phi_{\text{b}}-A_{\text{lat}}\Phi_{\text{lat}}-A_{\text{top}}\Phi_{\text{top}} \ .
\label{eq:evol}
\end{align}
With the substitution of eqs.~\eqref{eq:Gth}, \eqref{eq:Geta} and \eqref{eq:diff}, using the previously described dimensionless variables and numbers, and defining the dimensionless time $\tau=t/\mathcal{T}$, where $\mathcal{T}$ is a characteristic timescale, eq.~\eqref{eq:evol} can be rewritten as:
%
\begin{align}
\der{\DN{Bo}}{\tau}&=
-\left(\dfrac{\alpha_{\text{v}}\, \rho_{\text{v}}\, \mathcal{T}\, \DN{Ja}_{\text{v}}}{\rho_{\text{l}}\, \lambda_{\text{c}}^2}\right)\times \notag \\
&\qquad \left(\dfrac{\xi_{\text{b}}\, G_{\theta}}{H}+G_{\eta}
+\dfrac{\Pi_1^{\dagger}\, \DN{Sh}_0\, \Lambda_{\text{top}}}{4\, \DN{Ja}_{\text{v}}\, \xi_{\text{max}}}\right)\dfrac{1}{\DN{Bo}^{1/2}} \ ,
\label{eq:dimless}
\end{align}

Nevertheless, for dimensionless coherence and simplicity, we can make the first term on the right-hand side of the equation equal to one, which allows us to define the characteristic timescale as:
\begin{align}
\mathcal{T}&=\left(\dfrac{\rho_{\text{l}}\, \lambda_{\text{c}}^2}{\alpha_{\text{v}}\, \rho_{\text{v}}}\right)\dfrac{1}{\DN{Ja}_{\text{v}}} \ .
\label{eq:Tau}
\end{align}
Therefore, eq.~\eqref{eq:dimless} becomes:
\begin{align}
\der{\DN{Bo}}{\tau}&=-\left(\dfrac{\xi_{\text{b}}\, G_{\theta}}{H}+G_{\eta}+\dfrac{\Pi_1\, \DN{Sh}_0\, \Lambda_{\text{top}}}{\xi_{\text{max}}}\right)\dfrac{1}{\DN{Bo}^{1/2}} \ ,
\label{Bo:evol}
\end{align}
where we defined another dimensionless group as:
\begin{align}
\Pi_1&=\dfrac{\Pi_1^{\dagger}}{4\, \DN{Ja}_{\text{v}}} \ .
\label{eq:Pi3}
\end{align}
The dimensionless group $\Pi_1$ compares the mass diffusion from the drop towards the surrounding air with the liquid-vapor phase change below the drop.
In turn, the characteristic timescale $\mathcal{T}$ seems to be an evaporation timescale that depends on the plate temperature, by means of the Jakob number $\DN{Ja}$, and which considers a droplet evolution that does not include diffusion in air, by making $\Pi_1\rightarrow0$.

The evolution of the dimensionless size of the drop, represented by $\DN{Bo}$, is described by eq.~\eqref{Bo:evol}.
It is a nonlinear ODE, since all the terms in the paretheses are mostly nonlinear functions of $\DN{Bo}$, and, for the present study, its numerical solution is obtained using a Runge-Kutta-Fehlberg method.

\section{Parametric study}
\label{Sec:Param}

\begin{figure*}[t]
\centering
\includegraphics[width=0.65\textwidth]{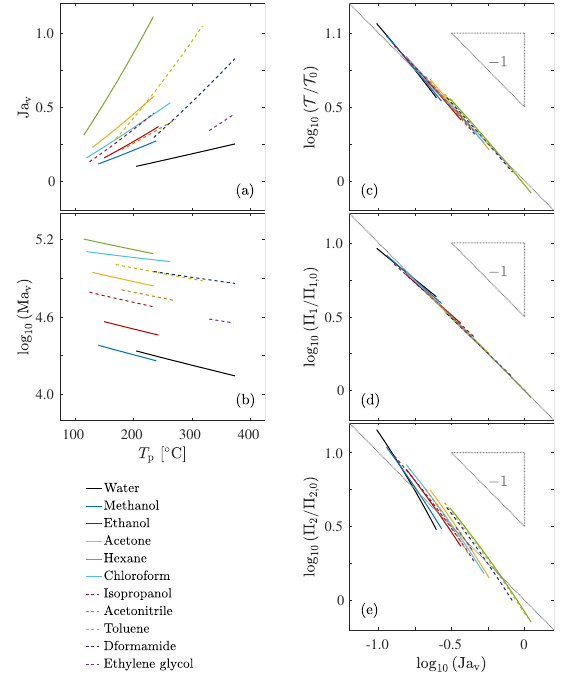}
\caption{
(a) Jacob number $\DN{Ja}_{\text{v}}$ and (b) Marangoni thermal number $\DN{Ma}_{\text{v}}$ as functions of the plate temperature $T_{\text{p}}$, within the range from the Leidenfrost temperature to the critical temperature, and considering the atmospheric pressure at Puebla, Mexico, at $\sim$2200 m above the sea level.
(c) Characteristic timescale $\mathcal{T}$ and dimensionless parameters (d) $\Pi_1$ and (e) $\Pi_2$, as functions of $\DN{Ja}_{\text{v}}$.
}
\label{fig:Param}
\end{figure*}

\begin{figure*}[t]
\centering
\includegraphics[width=0.75\textwidth]{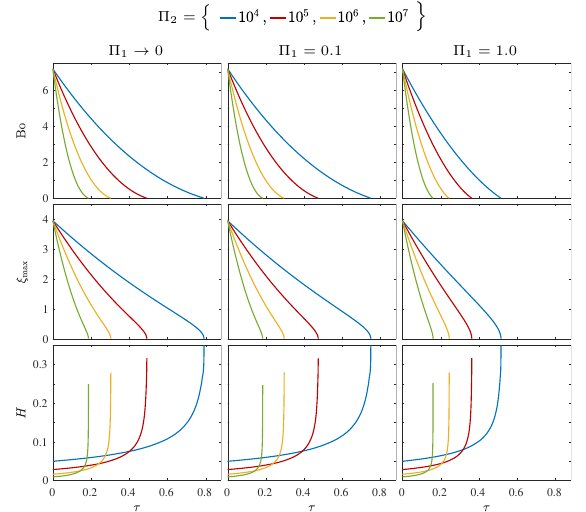}
\caption{
Evolution of the Bond number $\DN{Bo}$ (top row), the dimensionless maximum radius $\xi_{\text{max}}$ (middle row), and the dimensionless thickness of the vapor film $H$ (bottom row) as functions of the dimensionless time $\tau$, for different values of $\Pi_1$ (a given value for each column), several values of $\Pi_2$ (indicated with colors), and $\DN{Ma}_{\text{v}}/\DN{Ja}_{\text{v}}=0$ (neglecting the Marangoni convection flow).
}
\label{fig:Evol0}
\end{figure*}

\begin{figure*}[t]
\centering
\includegraphics[width=0.75\textwidth]{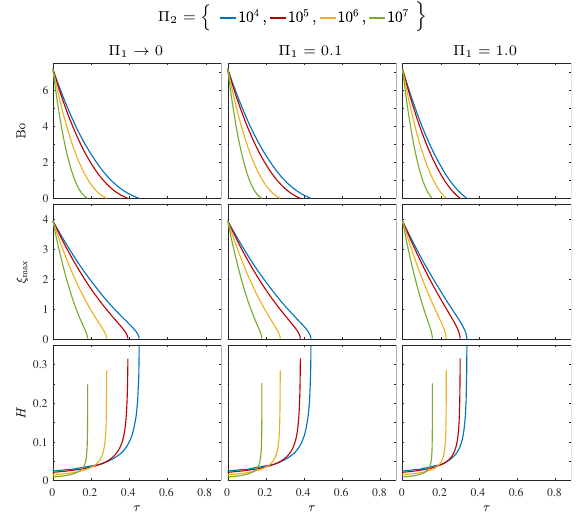}
\caption{
Evolution of the Bond number $\DN{Bo}$ (top row), the dimensionless maximum radius $\xi_{\text{max}}$ (middle row), and the dimensionless thickness of the vapor film $H$ (bottom row) as functions of the dimensionless time $\tau$, for different values of $\Pi_1$ (a given value for each column), several values of $\Pi_2$ (indicated with colors), and $\DN{Ma}_{\text{v}}/\DN{Ja}_{\text{v}}=10^5$ (taking into account the Marangoni convection flow).
}
\label{fig:Evol}
\end{figure*}

Temperature plays a major role in the determination of the characteristic timescale and the dimensionless parameters, since the thermophysical parameters of the vapor are calculated at the film temperature, which is taken as the average of the drop temperature $T_{\text{l}}$ and the plate temperature $T_{\text{p}}$.
The dependences of the Jacob number $\DN{Ja}_{\text{v}}$ and the thermal Marangoni number $\DN{Ma}_{\text{v}}$ on the plate temperature $T_{\text{p}}$, for several fluids, are shown in Fig.~\ref{fig:Param}.
For all the analized fluids, which properties had been calculated using the ideal gas approximation and empirical correlations~\cite{Poling,Yaws}, linear trends are observed between $\DN{Ja}_{\text{v}}$ and $T_{\text{p}}$, and between $\DN{Ma}_{\text{v}}$ and $T_{\text{p}}$.
Additionally, the characteristic timescale $\mathcal{T}$ and the dimensionless parameters $\Pi_1$ and $\Pi_2$ are also depicted in Fig.~\ref{fig:Param}, as functions of $\DN{Ja}_{\text{v}}$.
Taking average values of the thermophysical properties of the vapor (density $\overline{\rho}_{\text{v}}$, dynamic viscosity $\overline{\mu}_{\text{v}}$, thermal diffusivity $\overline{\alpha}_{\text{v}}$, and Spalding number $\overline{\DN{Bm}}$) we can define the following base parameters:
\begin{subequations}
\begin{align}
\mathcal{T}_0&=\dfrac{\rho_{\text{l}}\, \lambda_{\text{c}}^2}{\overline{\alpha}_{\text{v}}\, \overline{\rho}_{\text{v}}} \ , \\
\Pi_{1,0}&=\left(\dfrac{\overline{\rho}_{\text{air}}\, \mathcal{D}_{\text{v,air}}}{4\overline{\rho}_{\text{v}}\, \overline{\alpha}_{\text{v}}}\right)\, \ln\left(1+\overline{\DN{Bm}}\right) \ , \\
\Pi_{2,0}&
=\mathcal{T}_0 \left(\dfrac{\overline{\rho}_{\text{v}}\, g\, \lambda_{\text{c}}}{\overline{\mu}_{\text{v}}}\right) \ ,
\end{align}
\end{subequations}
allowing us to observe the deviation of the behavior of $\mathcal{T}$, $\Pi_1$ and $\Pi_2$ from the straightforward dependece of them on $\DN{Ja}_{\text{v}}$ described in eqs.~ \eqref{eq:Tau}, \eqref{eq:Pi2}, and \eqref{eq:Pi3}, respectively.
The averaged values are computed over the range in which the empirical correlations are valid~\cite{Yaws}, \emph{i.e.} between the corresponding Leidensfrost~\cite{Pacheco2021} and critical temperatures.
The values of $\mathcal{T}_0$, $\Pi_{1,0}$, and $\Pi_{2,0}$, for all the analized fluids, can be found  in SM~\cite{SuppMat}.

In Fig.~\ref{fig:Param}, we can observe that the timescale $\mathcal{T}$ and the dimensionless parameter $\Pi_1$, for all the fluids, are nearly proportional to $1/\DN{Ja}_{\text{v}}$, whereas $\Pi_2$ shows a clear deviation from the same trend (inversely proportional to $\DN{Ja}_{\text{v}}$).
The corresponding combinations of thermophysical parameters, that the three quantities involve, are strongly dependent of on the plate temperature $T_{\text{p}}$ and, as a consequence, on $\DN{Ja}_{\text{v}}$.
Nevertheless, the effect of $\DN{Ja}_{\text{v}}$ on $\mathcal{T}$ and $\Pi_1$ seems to counterbalance itself, not being the case for $\Pi_2$.
The employment of the base parameters $\mathcal{T}_0$, $\Pi_{1,0}$, and $\Pi_{2,0}$, taken as constant values for a given fluid, and considering an inverse proportionality with the Jacob number $\DN{Ja}_{\text{v}}$, can lead us to make a good estimation of the magnitude of the corresponding characteristic timescale $\mathcal{T}$ and the dimensionless parameters $\Pi_1$ and $\Pi_2$.
In the following steps of the proposed methodology, the dependences of $\mathcal{T}$, $\Pi_1$, and $\Pi_2$ on $T_{\text{p}}$, as given in eqs.~\eqref{eq:Pi1}, \eqref{eq:Pi2}, and \eqref{eq:Pi3}, are taken into account.

Remember that the dimensionless geometric properties of the drops that appear in eq.~\eqref{Bo:evol}, such as $\xi_{\text{b}}$, $\xi_{\text{max}}$, $\Lambda_{\text{top}}$, and $\DN{Sh}_0$, are functions of $\DN{Bo}$, as it has been depicted in Fig.~\ref{fig:shape} and eq.~\eqref{eq:Sh0}.
Also, the dimensionless quantities $G_{\theta}$, $G_{\eta}$ and $H$ depend on $\DN{Bo}$, $\Pi_2$, and $\DN{Ja}_{\text{v}}/\DN{Ma}_{\text{v}}$, as pointed out by eqs.~\eqref{eq:Gth}, \eqref{eq:Geta}, and \eqref{eq:HBo}.
Therefore, according to eq.~\eqref{Bo:evol}, the evolution in time $\tau$ of the Bond number $\DN{Bo}$ is driven by the values of the dimensionless groups $\Pi_1$, $\Pi_2$, and $\DN{Ja}_{\text{v}}/\DN{Ma}_{\text{v}}$.


Examples of the numerical solution of eq.~\eqref{Bo:evol} are presented in Figs.~\ref{fig:Evol0} and \ref{fig:Evol}, for different combinations of $\Pi_1$ and $\Pi_2$, along different orders of magnitude, and for $\DN{Ma}_{\text{v}}/\DN{Ja}_{\text{v}}=0$ and $\DN{Ma}_{\text{v}}/\DN{Ja}_{\text{v}}=10^5$, respectively.
According to the computed values of the physical parameters, for the studied fluids at temperatures between 200 $^{\circ}$C and 400 $^{\circ}$C, presented in SM~\cite{SuppMat}, the ranges are $\Pi_1\in\left[10^{-1},10^0\right]$, $\Pi_2\in\left[10^4,10^7\right]$, and $\DN{Ma}_{\text{v}}/\DN{Ja}_{\text{v}}\in\left[10^4,10^6\right]$.
The initial size of the drops is defined by $\DN{Bo}_0=7.185$, which corresponds to the maximum size for stable drops and vapor films~\cite{Snoeijer2009}, \emph{i.e.} drops with $r_{\text{max},0}=3.95\lambda_c$ or equivalently $\xi_{\text{max},0}=3.95$.

As time goes by, the size of the drops decreases due to the evaporation process.
This is observed in dimensionless terms as $\DN{Bo}$ shows a continuous contraction as the dimensionless time $\tau$ advances.
For the different combinations of $\Pi_1$ and $\Pi_2$, $\DN{Bo}$ performs a significant descent with the steepest slope at $\DN{Bo}_0$, becoming softer as the size of the drop reaches $\DN{Bo}\sim 2$, and reaching more moderate slopes as $\DN{Bo}$ approaches zero.
In turn, the dimensionless radius $\xi_{\text{max}}$ decreases with a nearly constant slope, until it reaches the size of the capillary length $\xi_{\text{max}}=1$, after which the slope becomes steeper as time goes by and the radius decreases.
The evolution of $\xi_{\text{max}}$ presents a vertical asymptote as the lifetime of the drop has been exhausted.
The ratio of film thickness to the bottom radius of the drop, $H$, starts at its smaller value at $\tau=0$, and increases monotonically as time elapses.
The slope of the evolution of $H$ in time $\tau$ grows, slowly up to 60\% of the drop lifetime, and then faster, tending to a vertical asymptote as the liquid is consumed and the drop is totally evaporated.
This is in agreement with the experimental findings observed in previous studies, in the $\DN{Bo}\ll1$ regime, where small drops suddenly take-off~\cite{Pomeau2012}.

The dimensionless groups $\Pi_1$, $\Pi_2$, and $\DN{Ma}_{\text{v}}/\DN{Ja}_{\text{v}}$ determine the precise evolution of a drop, keeping the aforementioned general characteristics of the dynamics of $\DN{Bo}$, $\xi_{\text{max}}$ and $H$.
To start, Fig.~\ref{fig:Evol0} corresponds to a situation for which the Marangoni convection flow is neglected, by taking $\DN{Ma}_{\text{v}}/\DN{Ja}_{\text{v}}=0$.
Consider a fixed value of $\Pi_1$, which corresponds to a single column of the results (all colors) plotted in Fig.~\ref{fig:Evol0}.
Large values of $\Pi_2$ imply the generation of a narrow vapor film with a relatively small $H$ at $\tau=0$, which entails a significant heat conduction throught the film and a large heat transfer rate toward the drop.
Therefore, fast evaporations and short drop lifetimes are observed for the large $\Pi_2$ cases.
Short values of $\Pi_2$ lead to the formation of a thick vapor film with relatively large $H$ at $t=0$, provoking a lessened heat conduction across the vapor and a reduced heat transfer rate toward the drop.
The consequence is a slower evaporation and a larger lifetime of drops with small values of $\Pi_2$.
Now, consider a fixed value of $\Pi_2$, which corresponds to the curves (all columns) with the same specific color presented in Fig.~\ref{fig:Evol0}.
For $\Pi_1\rightarrow0$, diffusion of the drop molecules into the surrounding air is negligible.
Consequently, the lifetime of a drop is enlarged to its maximum.
Values of $\Pi_1\leq 0.1$ provoke a slight reduction of the lifetime, whereas for $\Pi_1=1$ a significant shortening of the lifetime is observed.
The effect of $\Pi_1$ on the lifetime is more evident for smaller values of $\Pi_2$.
It is worthy to mention that $\Pi_1$ has no effect on the initial film thickness, thus $H$ at $t=0$ remains unchanged no matter the value of $\Pi_1$, for a given value of $\Pi_2$.

Fig.~\ref{fig:Evol} corresponds to $\DN{Ma}_{\text{v}}/\DN{Ja}_{\text{v}}=10^5$, for which the Marangoni convection flow is taken into account.
For $\Pi_2=10^4$, the smallest tested value of the dimensionless number, the effect of $\DN{Ma}_{\text{v}}/\DN{Ja}_{\text{v}}$ on the drop dynamics and the thickness of the vapor film is negligible, since the evolution of $\DN{Bo}$, $\xi_{\text{max}}$ and $H$, observed in Fig.~\ref{fig:Evol0} for $\DN{Ma}_{\text{v}}/\DN{Ja}_{\text{v}}=0$ and in Fig.~\ref{fig:Evol} for $\DN{Ma}_{\text{v}}/\DN{Ja}_{\text{v}}=10^5$, are almost identical for any value of $\Pi_1$.
For larger orders of magnitude of $\Pi_2$, the drop lifetime is reduced, comparing the curves in Fig.~\ref{fig:Evol} with respect to the ones in Fig.~\ref{fig:Evol0}, for the same value of $\Pi_1$.
The effect of $\DN{Ma}_{\text{v}}/\DN{Ja}_{\text{v}}$  increases as the value of $\Pi_2$ grows.

\section{Comparison with experiments}
\label{Sec:Exp}

\begin{figure*}[t]
    \centering
    \includegraphics[width=0.95\textwidth]{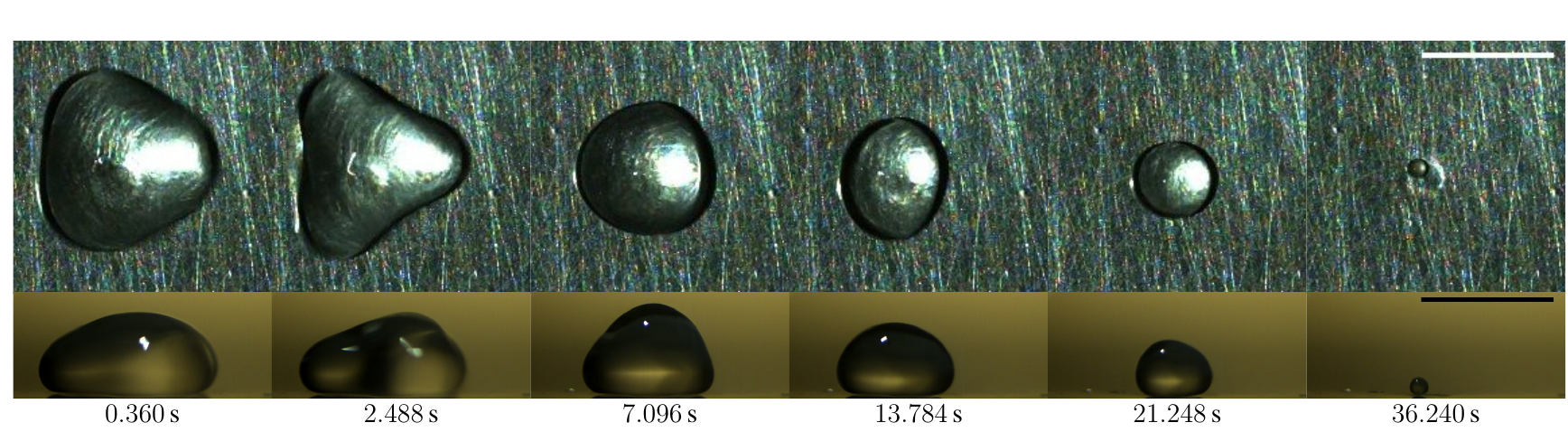}
    \caption{Image sequences representing the evolution of an isopropanol in Leidenfrost state, placed over an aluminum plate at  $T_{\text{p}}=300\, ^{\circ}\text{C}$.
    The top row presents the top view, whereas the bottom row shows the side view.
    Both scale bars correspond to 5 mm.}
    \label{fig:sequence}
\end{figure*}

\begin{figure*}[t]
    \centering
    \includegraphics[width=0.75\textwidth]{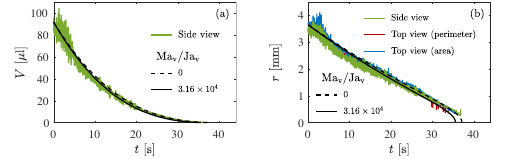}
    \caption{Evolution of the (a) volume and (b) radius of the isopropanol droplet in Leidenfrost state (shown in Fig.~\ref{fig:sequence}), as a function of time.
    In (a), the experimental volume was estimated using a solid of revolution (side view as in Fig.~\ref{fig:sequence}b, green curve), whereas in (b), the experimental radii were calculated from the maximum profile radius (side view as in Fig.~\ref{fig:sequence}b, green curve) or the projected perimeter and area (top view as in Fig.~\ref{fig:sequence}a, blue and red curves, respectively).
    The model trends (black continuous and dashed curves), in both (a) and (b),  were found using the numerical solution of eq.~\eqref{Bo:evol}, with $\DN{Bo}_0=3.513$, $\Pi_1=0.236$, $\Pi_2=2.113\times 10^5$, and the depicted values of $\DN{Ma}_{\text{v}}/\DN{Ja}_{\text{v}}$, respectively.
    The values of  $\mathcal{T}=142.92\, \text{s}$ and $\lambda_c=1.49\, \text{mm}$ were employed to calculate the volume, radii and time.}
    \label{fig:radii-comparison}
\end{figure*}

\begin{figure*}[t]
    \centering
    \includegraphics[width=1.0\textwidth]{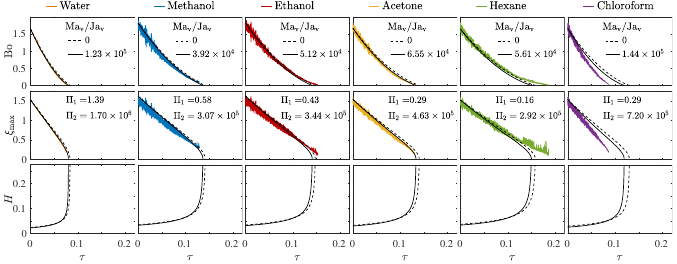}
    \caption{Evolution of the Bond number (top row), dimensionless maximum radius (middle row), and dimensionless film thickness (bottom row) of drops in Leidenfrost state, as a function of dimensionless time.
    Each column corresponds to a different fluid, water, methanol, ethanol, acetone, hexane and chloroform (from left to right).
    Color curves indicate experimental data, whereas black continuous curves and dashed curves depict the theoretical predictions for each fluid, considering and neglecting the Marangoni convection flow, respectively.
   The corresponding values of $\Pi_1$, $\Pi_2$, and $\DN{Ma}_{\text{v}}/\DN{Ja}_{\text{v}}$ are shown for each fluid (calculated in  SM~\cite{SuppMat}).
   The temperature of the aluminum plate is $T_{\text{p}}=250\, ^{\circ}\text{C}$, for all the fluids.}
    \label{fig:graphs}
\end{figure*}

Experiments were performed using a variation of the experimental setup as for a previous study~\cite{Pacheco2021}.
A 4 cm-thick polished aluminum substrate was placed on a hotplate, controlling the temperature with a closed loop (thermocouple and solid-state relay).
The substrate has a slight conical shape, with an angle of $\pi / 180$ radians measured from the horizontal, with the purpose of keeping the drops at a central position.
A drop was deposited on the aluminum plate, and two cameras were used to follow the evolution of the drop.
The cameras were placed such that their optical axes were orthogonal to each other, one recording the projection of the drop over the substrate (top view) and the other following the lateral projection (side view).
The side view was captured with a Photron SA3 high speed camera, with a frame rate of 125 fps, whereas the top view was obtained with a Photron UX100 high speed camera, with a frame rate of 50 fps.

In Fig.~\ref{fig:sequence}, the evolution of an isopropanol drop in the Leidenfrost state, on a plate at 300 $^{\circ}\text{C}$, is presented as a sequence of images from different instants.
From the top and side views, image processing allowed us to identify the top and lateral profiles of the drop at each frame, respectively.
Taking the average lateral profile, from the left and right parts, led to the direct calculation of the maximum radius $r_{\text{max}}$ and, considering it as the generatrix of a solid of revolution, to the estimation of the volume $V$.
From the top profile, the projected perimeter and area of the drop were recovered, which also allowed us to compute two additional values of $r_{\text{max}}$, considering equivalent circles of the same perimeter and area, respectively.
Examples of the aforementioned experimental results are depicted in Fig.~\ref{fig:radii-comparison}, corresponding to the Isopropanol drop presented in Fig.~\ref{fig:sequence}.
In Fig.~\ref{fig:radii-comparison}a, the temporal evolution of the volume $V$ is compared with its theoretical behavior, obtained from the solution of eq.~\eqref{Bo:evol}, for the equivalent dimensionless parameters.
The theoretical curve accurately describes the trend that the experimentally measured volume follows.
In Fig.~\ref{fig:radii-comparison}b, the three values of maximum radius $r_{\text{max}}$, obtained from the application of the different procedures to the top and side views,  are depicted as a function of time and compared to the corresponding result, yielded from eq.~\eqref{Bo:evol}.
The experimental data, recovered from the equivalent circles of the projected perimeter and area, and directly from the lateral profile, show equivalent behaviours with almost indistinguishable trends.
Once more, the theoretical results reliably represent the experimental findings, indicating the usefulness and validity of eq.~\eqref{Bo:evol}.

In order to evaluate the generality of the theoretical approach, experiments with different fluids were performed.
Also, in view of the coherence between the experimental data, we decided to present in the following, only the results (volume and maximum radius) obtained from the side view.
The initial time $t=0$ of each experiment has been considered as any time after the moment of drop deposition on the aluminum plate.
For comparison, a volume $V_0\approx 7.7 \lambda_c^3$ was taken as the initial volume at $t=0$ for all the drops, corresponding to a maximum radius of $r_{\text{max},0}\approx 1.5 \lambda_c$, which is below the upper limit to produce drops in the Leidenfrost state with stable vapor cushions~\cite{Snoeijer2009}.
The results for the different tested fluids are presented in Fig.~\ref{fig:graphs}.
For water, methanol, ethanol and acetone drops, good agreement is observed between the theoretical predictions and the measured evolutions of the drops.
In contrast, for the case of hexane, the dynamics is properly followed by the theoretical approach, while the drop is relatively large $r_{\text{max}}>0.5\lambda_c$, after which, the evaporation is overestimated and the predicted lifetime is slightly smaller than the experimentally observed time.
When hexane droplets reach the abovementioned size, they start bouncing on the plate, exhibiting a dynamic behavior and changing drastically the vapor film thickness~\cite{Pomeau2012,Chen2014,Sun2019,Graeber2021}.
This behavior diverges from the quasi-static assumption, upon which the presented model is based.
Also, the trend that is displayed by the experimental data for chloroform is only followed qualitatively by the theoretical prediction of the model, since the evaporation is clearly underestimated, hence leading to larger predicted lifetimes than the observed during the experiments.
Chloroform drops exhibit spontaneous axial rotation around the $z$-axis, a situation that modifies the velocity boundary condition at the drop-vapor film interface~\cite{Yang2023}.
As a consequence, the vapor moves both in the angular and radial direction in the film, which accelerates its drainage, shrinking the thickness of the vapor film, and increasing the evaporation rate of the drop.

\section{Conclusions}
\label{Sec:Conclusions}


A simple model to follow the complete evolution of a sessile drop in Leidenfrost state has been proposed.
A non-linear ODE, in terms of the instantaneous Bond number, has been obtained to calculate the size evolution of the drop over time.
The equation includes the following features:
\begin{enumerate}
\item The shape of a sessile drop is described by solving the Young-Laplace equation.
\item The velocity, pressure and temperature fields of the vapor film, between the drop and the hot plate, are recovered by means of the application of Hankel transforms, which is valid for any drop size and vapor film thickness, below their stability threshold.
\item An estimation of the effect of the thermo-capillary Marangoni convection flow within the drop is made, which is taken as a boundary condition for the velocity field at the vapor film, without a complete numerical simulation of the drop internal flow. 
\item The evaporation is considered at the entire surface of the drop, beneath the drop, due to the heat transfer across the vapor film, sideways, due to the proximity of the hot plate, and above the drop, due to diffusion in the surrounding air.
\item The weight of the drop is balanced against the pressure exerted by the vapor film.
\end{enumerate}
Not a single fitting parameter was required to develop the proposed methodology, since the properties of the fluids were calculated using the ideal gas approximation and empirical correlations~\cite{Poling,Yaws}, according to the corresponding temperatures, \emph{i.e.} $T_{\text{l}}$ for the liquid
phase and $\left(T_{\text{p}}+T_{\text{l}}\right)/2$ for the phase at the vapor film.
It is also noteworthy to mention that, since our model does not employs the lubrication approximation, it is valid for any size of drops and their corresponding film thickness.

Our model describes with accuracy, the dynamics of the drop size, yielding to a good estimate of the drop lifetime and the vapor film thickness.
Several organic and inorganic compounds, which exact composition and molecular structure is very different, had been used to test the validity of the model.
Good agreement with most of the tested fluids (water, isopropanol, methanol, ethanol, and acetone) has been observed, between the experimental results and the model predictions.
Nevertheless, when other phenomena occurs, our model overestimates (for hexane drops that bounce) or underestimates (for chloroform that rotate) the evaporation rate.
When dynamic phenomena, such as bouncing or spontaneous rotation is not observed, our model provides a simple, accurate and efficient tool to describe the evolution and estimate the lifetime and vapor film thickness for drops in the Leidenfrost state.



% ...



\bibliography{Leidenfrost_V07}

\begin{thebibliography}{56}%
\makeatletter
\providecommand \@ifxundefined [1]{%
 \@ifx{#1\undefined}
}%
\providecommand \@ifnum [1]{%
 \ifnum #1\expandafter \@firstoftwo
 \else \expandafter \@secondoftwo
 \fi
}%
\providecommand \@ifx [1]{%
 \ifx #1\expandafter \@firstoftwo
 \else \expandafter \@secondoftwo
 \fi
}%
\providecommand \natexlab [1]{#1}%
\providecommand \enquote  [1]{``#1''}%
\providecommand \bibnamefont  [1]{#1}%
\providecommand \bibfnamefont [1]{#1}%
\providecommand \citenamefont [1]{#1}%
\providecommand \href@noop [0]{\@secondoftwo}%
\providecommand \href [0]{\begingroup \@sanitize@url \@href}%
\providecommand \@href[1]{\@@startlink{#1}\@@href}%
\providecommand \@@href[1]{\endgroup#1\@@endlink}%
\providecommand \@sanitize@url [0]{\catcode `\\12\catcode `\$12\catcode
  `\&12\catcode `\#12\catcode `\^12\catcode `\_12\catcode `\%12\relax}%
\providecommand \@@startlink[1]{}%
\providecommand \@@endlink[0]{}%
\providecommand \url  [0]{\begingroup\@sanitize@url \@url }%
\providecommand \@url [1]{\endgroup\@href {#1}{\urlprefix }}%
\providecommand \urlprefix  [0]{URL }%
\providecommand \Eprint [0]{\href }%
\providecommand \doibase [0]{https://doi.org/}%
\providecommand \selectlanguage [0]{\@gobble}%
\providecommand \bibinfo  [0]{\@secondoftwo}%
\providecommand \bibfield  [0]{\@secondoftwo}%
\providecommand \translation [1]{[#1]}%
\providecommand \BibitemOpen [0]{}%
\providecommand \bibitemStop [0]{}%
\providecommand \bibitemNoStop [0]{.\EOS\space}%
\providecommand \EOS [0]{\spacefactor3000\relax}%
\providecommand \BibitemShut  [1]{\csname bibitem#1\endcsname}%
\let\auto@bib@innerbib\@empty
\bibitem [{\citenamefont {Leidenfrost}(1756)}]{Leidenfrost1756}%
  \BibitemOpen
  \bibfield  {author} {\bibinfo {author} {\bibfnamefont {J.}~\bibnamefont
  {Leidenfrost}},\ }\href@noop {} {\bibinfo {title} {De aquae communis
  nonnullis qualitatibus tractatus}} (\bibinfo {year} {1756})\BibitemShut
  {NoStop}%
\bibitem [{\citenamefont {Quere}(2013)}]{Quere2013}%
  \BibitemOpen
  \bibfield  {author} {\bibinfo {author} {\bibfnamefont {D.}~\bibnamefont
  {Quere}},\ }\bibfield  {title} {\bibinfo {title} {Leidenfrost dynamics},\
  }\href@noop {} {\bibfield  {journal} {\bibinfo  {journal} {Annu. Rev. Fluid
  Mech.}\ }\textbf {\bibinfo {volume} {45}},\ \bibinfo {pages} {197} (\bibinfo
  {year} {2013})}\BibitemShut {NoStop}%
\bibitem [{\citenamefont {Ajaev}\ and\ \citenamefont
  {Kabov}(2021)}]{Kabov2021}%
  \BibitemOpen
  \bibfield  {author} {\bibinfo {author} {\bibfnamefont {V.}~\bibnamefont
  {Ajaev}}\ and\ \bibinfo {author} {\bibfnamefont {O.}~\bibnamefont {Kabov}},\
  }\bibfield  {title} {\bibinfo {title} {Levitation and self-organization of
  droplets},\ }\href@noop {} {\bibfield  {journal} {\bibinfo  {journal} {Annu.
  Rev. Fluid Mech.}\ }\textbf {\bibinfo {volume} {53}},\ \bibinfo {pages} {203}
  (\bibinfo {year} {2021})}\BibitemShut {NoStop}%
\bibitem [{\citenamefont {Stewart}(2022)}]{Stewart2022}%
  \BibitemOpen
  \bibfield  {author} {\bibinfo {author} {\bibfnamefont {S.}~\bibnamefont
  {Stewart}},\ }\bibfield  {title} {\bibinfo {title} {Leidenfrost drop
  dynamics: a forgotten past and modern day rediscoveries},\ }\href@noop {}
  {\bibfield  {journal} {\bibinfo  {journal} {Eur. J. Phys.}\ }\textbf
  {\bibinfo {volume} {43}},\ \bibinfo {pages} {1} (\bibinfo {year}
  {2022})}\BibitemShut {NoStop}%
\bibitem [{\citenamefont {Linke}\ \emph {et~al.}(2006)\citenamefont {Linke},
  \citenamefont {Aleman}, \citenamefont {Melling}, \citenamefont {Taormina},
  \citenamefont {Francis}, \citenamefont {Dow-Hygelund}, \citenamefont
  {Narayanan}, \citenamefont {Taylor},\ and\ \citenamefont
  {Stout}}]{Linke2006}%
  \BibitemOpen
  \bibfield  {author} {\bibinfo {author} {\bibfnamefont {H.}~\bibnamefont
  {Linke}}, \bibinfo {author} {\bibfnamefont {B.~J.}\ \bibnamefont {Aleman}},
  \bibinfo {author} {\bibfnamefont {L.~D.}\ \bibnamefont {Melling}}, \bibinfo
  {author} {\bibfnamefont {M.~J.}\ \bibnamefont {Taormina}}, \bibinfo {author}
  {\bibfnamefont {M.~J.}\ \bibnamefont {Francis}}, \bibinfo {author}
  {\bibfnamefont {C.~C.}\ \bibnamefont {Dow-Hygelund}}, \bibinfo {author}
  {\bibfnamefont {V.}~\bibnamefont {Narayanan}}, \bibinfo {author}
  {\bibfnamefont {R.~P.}\ \bibnamefont {Taylor}},\ and\ \bibinfo {author}
  {\bibfnamefont {A.}~\bibnamefont {Stout}},\ }\bibfield  {title} {\bibinfo
  {title} {Self-propelled leidenfrost droplets},\ }\href@noop {} {\bibfield
  {journal} {\bibinfo  {journal} {Phys. Rev. Lett.}\ }\textbf {\bibinfo
  {volume} {96}},\ \bibinfo {pages} {1} (\bibinfo {year} {2006})}\BibitemShut
  {NoStop}%
\bibitem [{\citenamefont {Lagubeau}\ \emph {et~al.}(2011)\citenamefont
  {Lagubeau}, \citenamefont {Le~Merrer}, \citenamefont {Clanet},\ and\
  \citenamefont {Quere}}]{Quere2011}%
  \BibitemOpen
  \bibfield  {author} {\bibinfo {author} {\bibfnamefont {G.}~\bibnamefont
  {Lagubeau}}, \bibinfo {author} {\bibfnamefont {M.}~\bibnamefont {Le~Merrer}},
  \bibinfo {author} {\bibfnamefont {C.}~\bibnamefont {Clanet}},\ and\ \bibinfo
  {author} {\bibfnamefont {D.}~\bibnamefont {Quere}},\ }\bibfield  {title}
  {\bibinfo {title} {Leidenfrost on a ratchet},\ }\href@noop {} {\bibfield
  {journal} {\bibinfo  {journal} {Nature Phys.}\ }\textbf {\bibinfo {volume}
  {7}},\ \bibinfo {pages} {395} (\bibinfo {year} {2011})}\BibitemShut {NoStop}%
\bibitem [{\citenamefont {Cousins}\ \emph {et~al.}(2012)\citenamefont
  {Cousins}, \citenamefont {Goldstein}, \citenamefont {Jaworski},\ and\
  \citenamefont {Pesci}}]{Cousins2012}%
  \BibitemOpen
  \bibfield  {author} {\bibinfo {author} {\bibfnamefont {T.}~\bibnamefont
  {Cousins}}, \bibinfo {author} {\bibfnamefont {R.}~\bibnamefont {Goldstein}},
  \bibinfo {author} {\bibfnamefont {J.}~\bibnamefont {Jaworski}},\ and\
  \bibinfo {author} {\bibfnamefont {A.}~\bibnamefont {Pesci}},\ }\bibfield
  {title} {\bibinfo {title} {A ratchet trap for leidenfrost drops},\
  }\href@noop {} {\bibfield  {journal} {\bibinfo  {journal} {J. Fluid Mech.}\
  }\textbf {\bibinfo {volume} {696}},\ \bibinfo {pages} {215} (\bibinfo {year}
  {2012})}\BibitemShut {NoStop}%
\bibitem [{\citenamefont {Sobac}\ \emph {et~al.}(2017)\citenamefont {Sobac},
  \citenamefont {Rednikov}, \citenamefont {Dorbolo},\ and\ \citenamefont
  {Colinet}}]{Colinet2017}%
  \BibitemOpen
  \bibfield  {author} {\bibinfo {author} {\bibfnamefont {B.}~\bibnamefont
  {Sobac}}, \bibinfo {author} {\bibfnamefont {A.}~\bibnamefont {Rednikov}},
  \bibinfo {author} {\bibfnamefont {S.}~\bibnamefont {Dorbolo}},\ and\ \bibinfo
  {author} {\bibfnamefont {P.}~\bibnamefont {Colinet}},\ }\bibfield  {title}
  {\bibinfo {title} {Self-propelled leidenfrost drops on a thermal gradient: A
  theoretical study},\ }\href@noop {} {\bibfield  {journal} {\bibinfo
  {journal} {Phys. Fluids}\ }\textbf {\bibinfo {volume} {29}},\ \bibinfo
  {pages} {1} (\bibinfo {year} {2017})}\BibitemShut {NoStop}%
\bibitem [{\citenamefont {Bouillant}\ \emph {et~al.}(2018)\citenamefont
  {Bouillant}, \citenamefont {Mouterde}, \citenamefont {Bourrianne},
  \citenamefont {Lagarde}, \citenamefont {Clanet},\ and\ \citenamefont
  {Quere}}]{Quere2018}%
  \BibitemOpen
  \bibfield  {author} {\bibinfo {author} {\bibfnamefont {A.}~\bibnamefont
  {Bouillant}}, \bibinfo {author} {\bibfnamefont {T.}~\bibnamefont {Mouterde}},
  \bibinfo {author} {\bibfnamefont {P.}~\bibnamefont {Bourrianne}}, \bibinfo
  {author} {\bibfnamefont {A.}~\bibnamefont {Lagarde}}, \bibinfo {author}
  {\bibfnamefont {C.}~\bibnamefont {Clanet}},\ and\ \bibinfo {author}
  {\bibfnamefont {D.}~\bibnamefont {Quere}},\ }\bibfield  {title} {\bibinfo
  {title} {Leidenfrost wheels},\ }\href@noop {} {\bibfield  {journal} {\bibinfo
   {journal} {Nature Phys.}\ }\textbf {\bibinfo {volume} {14}},\ \bibinfo
  {pages} {1188} (\bibinfo {year} {2018})}\BibitemShut {NoStop}%
\bibitem [{\citenamefont {Gauthier}\ \emph {et~al.}(2019)\citenamefont
  {Gauthier}, \citenamefont {Diddens}, \citenamefont {Proville}, \citenamefont
  {Lohse},\ and\ \citenamefont {van~der Meer}}]{Lohse2019}%
  \BibitemOpen
  \bibfield  {author} {\bibinfo {author} {\bibfnamefont {A.}~\bibnamefont
  {Gauthier}}, \bibinfo {author} {\bibfnamefont {C.}~\bibnamefont {Diddens}},
  \bibinfo {author} {\bibfnamefont {R.}~\bibnamefont {Proville}}, \bibinfo
  {author} {\bibfnamefont {D.}~\bibnamefont {Lohse}},\ and\ \bibinfo {author}
  {\bibfnamefont {D.}~\bibnamefont {van~der Meer}},\ }\bibfield  {title}
  {\bibinfo {title} {Self-propulsion of inverse leidenfrost drops on a
  cryogenic bath},\ }\href@noop {} {\bibfield  {journal} {\bibinfo  {journal}
  {Proc. Natl. Acad. Sci.}\ }\textbf {\bibinfo {volume} {116}},\ \bibinfo
  {pages} {1174} (\bibinfo {year} {2019})}\BibitemShut {NoStop}%
\bibitem [{\citenamefont {Matsumoto}\ and\ \citenamefont
  {Hasegawa}(2021)}]{Matsumoto2021}%
  \BibitemOpen
  \bibfield  {author} {\bibinfo {author} {\bibfnamefont {R.}~\bibnamefont
  {Matsumoto}}\ and\ \bibinfo {author} {\bibfnamefont {K.}~\bibnamefont
  {Hasegawa}},\ }\bibfield  {title} {\bibinfo {title} {Self-propelled
  leidenfrost droplets on a heated glycerol pool},\ }\href@noop {} {\bibfield
  {journal} {\bibinfo  {journal} {Sci. Rep.}\ }\textbf {\bibinfo {volume}
  {11}},\ \bibinfo {pages} {1} (\bibinfo {year} {2021})}\BibitemShut {NoStop}%
\bibitem [{\citenamefont {Celestini}\ \emph {et~al.}(2012)\citenamefont
  {Celestini}, \citenamefont {Frisch},\ and\ \citenamefont
  {Pomeau}}]{Pomeau2012}%
  \BibitemOpen
  \bibfield  {author} {\bibinfo {author} {\bibfnamefont {F.}~\bibnamefont
  {Celestini}}, \bibinfo {author} {\bibfnamefont {T.}~\bibnamefont {Frisch}},\
  and\ \bibinfo {author} {\bibfnamefont {Y.}~\bibnamefont {Pomeau}},\
  }\bibfield  {title} {\bibinfo {title} {Take off of small leidenfrost
  droplets},\ }\href@noop {} {\bibfield  {journal} {\bibinfo  {journal} {Phys.
  Rev. Lett.}\ }\textbf {\bibinfo {volume} {109}},\ \bibinfo {pages} {1}
  (\bibinfo {year} {2012})}\BibitemShut {NoStop}%
\bibitem [{\citenamefont {Liu}\ \emph {et~al.}(2014)\citenamefont {Liu},
  \citenamefont {Ghigliotti}, \citenamefont {Feng},\ and\ \citenamefont
  {Chen}}]{Chen2014}%
  \BibitemOpen
  \bibfield  {author} {\bibinfo {author} {\bibfnamefont {F.}~\bibnamefont
  {Liu}}, \bibinfo {author} {\bibfnamefont {G.}~\bibnamefont {Ghigliotti}},
  \bibinfo {author} {\bibfnamefont {J.}~\bibnamefont {Feng}},\ and\ \bibinfo
  {author} {\bibfnamefont {C.-H.}\ \bibnamefont {Chen}},\ }\bibfield  {title}
  {\bibinfo {title} {Self-propelled jumping upon drop coalescence on
  leidenfrost surfaces},\ }\href@noop {} {\bibfield  {journal} {\bibinfo
  {journal} {J. Fluid Mech.}\ }\textbf {\bibinfo {volume} {752}},\ \bibinfo
  {pages} {22} (\bibinfo {year} {2014})}\BibitemShut {NoStop}%
\bibitem [{\citenamefont {Lyu}\ \emph {et~al.}(2019)\citenamefont {Lyu},
  \citenamefont {Mathai}, \citenamefont {Wang}, \citenamefont {Sobac},
  \citenamefont {Colinet}, \citenamefont {Lohse},\ and\ \citenamefont
  {Sun}}]{Sun2019}%
  \BibitemOpen
  \bibfield  {author} {\bibinfo {author} {\bibfnamefont {S.}~\bibnamefont
  {Lyu}}, \bibinfo {author} {\bibfnamefont {V.}~\bibnamefont {Mathai}},
  \bibinfo {author} {\bibfnamefont {Y.}~\bibnamefont {Wang}}, \bibinfo {author}
  {\bibfnamefont {B.}~\bibnamefont {Sobac}}, \bibinfo {author} {\bibfnamefont
  {P.}~\bibnamefont {Colinet}}, \bibinfo {author} {\bibfnamefont
  {D.}~\bibnamefont {Lohse}},\ and\ \bibinfo {author} {\bibfnamefont
  {C.}~\bibnamefont {Sun}},\ }\bibfield  {title} {\bibinfo {title} {Final fate
  of a leidenfrost droplet: Explosion or takeoff},\ }\href@noop {} {\bibfield
  {journal} {\bibinfo  {journal} {Sci. Adv.}\ }\textbf {\bibinfo {volume}
  {5}},\ \bibinfo {pages} {1} (\bibinfo {year} {2019})}\BibitemShut {NoStop}%
\bibitem [{\citenamefont {Graeber}\ \emph {et~al.}(2021)\citenamefont
  {Graeber}, \citenamefont {Regulagadda}, \citenamefont {Hodel}, \citenamefont
  {Kuttel}, \citenamefont {Landolf}, \citenamefont {Schutzius},\ and\
  \citenamefont {Poulikakos}}]{Graeber2021}%
  \BibitemOpen
  \bibfield  {author} {\bibinfo {author} {\bibfnamefont {G.}~\bibnamefont
  {Graeber}}, \bibinfo {author} {\bibfnamefont {K.}~\bibnamefont
  {Regulagadda}}, \bibinfo {author} {\bibfnamefont {P.}~\bibnamefont {Hodel}},
  \bibinfo {author} {\bibfnamefont {C.}~\bibnamefont {Kuttel}}, \bibinfo
  {author} {\bibfnamefont {D.}~\bibnamefont {Landolf}}, \bibinfo {author}
  {\bibfnamefont {T.}~\bibnamefont {Schutzius}},\ and\ \bibinfo {author}
  {\bibfnamefont {D.}~\bibnamefont {Poulikakos}},\ }\bibfield  {title}
  {\bibinfo {title} {Leidenfrost droplet trampolining},\ }\href@noop {}
  {\bibfield  {journal} {\bibinfo  {journal} {Nat. Commun.}\ }\textbf {\bibinfo
  {volume} {12}},\ \bibinfo {pages} {1} (\bibinfo {year} {2021})}\BibitemShut
  {NoStop}%
\bibitem [{\citenamefont {Caswell}(2014)}]{Caswell2014}%
  \BibitemOpen
  \bibfield  {author} {\bibinfo {author} {\bibfnamefont {T.}~\bibnamefont
  {Caswell}},\ }\bibfield  {title} {\bibinfo {title} {Dynamics of the vapor
  layer below a leidenfrost drop},\ }\href@noop {} {\bibfield  {journal}
  {\bibinfo  {journal} {Phys. Rev. E}\ }\textbf {\bibinfo {volume} {90}},\
  \bibinfo {pages} {1} (\bibinfo {year} {2014})}\BibitemShut {NoStop}%
\bibitem [{\citenamefont {Celestini}\ \emph {et~al.}(2014)\citenamefont
  {Celestini}, \citenamefont {Frisch}, \citenamefont {Cohen}, \citenamefont
  {Raufaste}, \citenamefont {Duchemin},\ and\ \citenamefont
  {Pomeau}}]{Pomeau2014}%
  \BibitemOpen
  \bibfield  {author} {\bibinfo {author} {\bibfnamefont {F.}~\bibnamefont
  {Celestini}}, \bibinfo {author} {\bibfnamefont {T.}~\bibnamefont {Frisch}},
  \bibinfo {author} {\bibfnamefont {A.}~\bibnamefont {Cohen}}, \bibinfo
  {author} {\bibfnamefont {C.}~\bibnamefont {Raufaste}}, \bibinfo {author}
  {\bibfnamefont {L.}~\bibnamefont {Duchemin}},\ and\ \bibinfo {author}
  {\bibfnamefont {Y.}~\bibnamefont {Pomeau}},\ }\bibfield  {title} {\bibinfo
  {title} {Two dimensional leidenfrost droplets in a hele-shaw cell},\
  }\href@noop {} {\bibfield  {journal} {\bibinfo  {journal} {Phys. Fluids}\
  }\textbf {\bibinfo {volume} {26}},\ \bibinfo {pages} {1} (\bibinfo {year}
  {2014})}\BibitemShut {NoStop}%
\bibitem [{\citenamefont {Ma}\ \emph {et~al.}(2017)\citenamefont {Ma},
  \citenamefont {Lietor-Santos},\ and\ \citenamefont {Burton}}]{Burton2017}%
  \BibitemOpen
  \bibfield  {author} {\bibinfo {author} {\bibfnamefont {X.}~\bibnamefont
  {Ma}}, \bibinfo {author} {\bibfnamefont {J.-J.}\ \bibnamefont
  {Lietor-Santos}},\ and\ \bibinfo {author} {\bibfnamefont {J.}~\bibnamefont
  {Burton}},\ }\bibfield  {title} {\bibinfo {title} {Star-shaped oscillations
  of leidenfrost drops},\ }\href@noop {} {\bibfield  {journal} {\bibinfo
  {journal} {Phys. Rev. Fluids}\ }\textbf {\bibinfo {volume} {2}},\ \bibinfo
  {pages} {1} (\bibinfo {year} {2017})}\BibitemShut {NoStop}%
\bibitem [{\citenamefont {Ma}\ and\ \citenamefont {Burton}(2018)}]{Burton2018}%
  \BibitemOpen
  \bibfield  {author} {\bibinfo {author} {\bibfnamefont {X.}~\bibnamefont
  {Ma}}\ and\ \bibinfo {author} {\bibfnamefont {J.}~\bibnamefont {Burton}},\
  }\bibfield  {title} {\bibinfo {title} {Self-organized oscillations of
  leidenfrost drops},\ }\href@noop {} {\bibfield  {journal} {\bibinfo
  {journal} {J. Fluid Mech.}\ }\textbf {\bibinfo {volume} {846}},\ \bibinfo
  {pages} {263} (\bibinfo {year} {2018})}\BibitemShut {NoStop}%
\bibitem [{\citenamefont {Bouillant}\ \emph {et~al.}(2021)\citenamefont
  {Bouillant}, \citenamefont {Cohen}, \citenamefont {Clanet},\ and\
  \citenamefont {Quere}}]{Quere2021}%
  \BibitemOpen
  \bibfield  {author} {\bibinfo {author} {\bibfnamefont {A.}~\bibnamefont
  {Bouillant}}, \bibinfo {author} {\bibfnamefont {C.}~\bibnamefont {Cohen}},
  \bibinfo {author} {\bibfnamefont {C.}~\bibnamefont {Clanet}},\ and\ \bibinfo
  {author} {\bibfnamefont {D.}~\bibnamefont {Quere}},\ }\bibfield  {title}
  {\bibinfo {title} {Self-excitation of leidenfrost drops and consequences on
  their stability},\ }\href@noop {} {\bibfield  {journal} {\bibinfo  {journal}
  {Proc. Natl. Acad. Sci.}\ }\textbf {\bibinfo {volume} {118}},\ \bibinfo
  {pages} {1} (\bibinfo {year} {2021})}\BibitemShut {NoStop}%
\bibitem [{\citenamefont {Raufaste}\ \emph {et~al.}(2016)\citenamefont
  {Raufaste}, \citenamefont {Bouret},\ and\ \citenamefont
  {Celestini}}]{Raufaste2016}%
  \BibitemOpen
  \bibfield  {author} {\bibinfo {author} {\bibfnamefont {C.}~\bibnamefont
  {Raufaste}}, \bibinfo {author} {\bibfnamefont {Y.}~\bibnamefont {Bouret}},\
  and\ \bibinfo {author} {\bibfnamefont {F.}~\bibnamefont {Celestini}},\
  }\bibfield  {title} {\bibinfo {title} {Reactive leidenfrost droplets},\
  }\href@noop {} {\bibfield  {journal} {\bibinfo  {journal} {Eur. Phys. Lett.}\
  }\textbf {\bibinfo {volume} {114}},\ \bibinfo {pages} {1} (\bibinfo {year}
  {2016})}\BibitemShut {NoStop}%
\bibitem [{\citenamefont {Pacheco-Vazquez}\ \emph {et~al.}(2021)\citenamefont
  {Pacheco-Vazquez}, \citenamefont {Ledesma-Alonso}, \citenamefont
  {Palacio-Rangel},\ and\ \citenamefont {Moreau}}]{Pacheco2021}%
  \BibitemOpen
  \bibfield  {author} {\bibinfo {author} {\bibfnamefont {F.}~\bibnamefont
  {Pacheco-Vazquez}}, \bibinfo {author} {\bibfnamefont {R.}~\bibnamefont
  {Ledesma-Alonso}}, \bibinfo {author} {\bibfnamefont {J.}~\bibnamefont
  {Palacio-Rangel}},\ and\ \bibinfo {author} {\bibfnamefont {F.}~\bibnamefont
  {Moreau}},\ }\bibfield  {title} {\bibinfo {title} {Triple leidenfrost effect:
  Preventing coalescence of drops on a hot plate},\ }\href@noop {} {\bibfield
  {journal} {\bibinfo  {journal} {Phys. Rev. Lett.}\ }\textbf {\bibinfo
  {volume} {127}},\ \bibinfo {pages} {1} (\bibinfo {year} {2021})}\BibitemShut
  {NoStop}%
\bibitem [{\citenamefont {Lyu}\ \emph {et~al.}(2021)\citenamefont {Lyu},
  \citenamefont {Tan}, \citenamefont {Wakata}, \citenamefont {Yang},
  \citenamefont {Law}, \citenamefont {Lohse},\ and\ \citenamefont
  {Sun}}]{Sun2021}%
  \BibitemOpen
  \bibfield  {author} {\bibinfo {author} {\bibfnamefont {S.}~\bibnamefont
  {Lyu}}, \bibinfo {author} {\bibfnamefont {H.}~\bibnamefont {Tan}}, \bibinfo
  {author} {\bibfnamefont {Y.}~\bibnamefont {Wakata}}, \bibinfo {author}
  {\bibfnamefont {X.}~\bibnamefont {Yang}}, \bibinfo {author} {\bibfnamefont
  {C.}~\bibnamefont {Law}}, \bibinfo {author} {\bibfnamefont {D.}~\bibnamefont
  {Lohse}},\ and\ \bibinfo {author} {\bibfnamefont {C.}~\bibnamefont {Sun}},\
  }\bibfield  {title} {\bibinfo {title} {On explosive boiling of a
  multicomponent leidenfrost drop},\ }\href@noop {} {\bibfield  {journal}
  {\bibinfo  {journal} {Proc. Natl. Acad. Sci.}\ }\textbf {\bibinfo {volume}
  {118}},\ \bibinfo {pages} {1} (\bibinfo {year} {2021})}\BibitemShut {NoStop}%
\bibitem [{\citenamefont {Xu}\ and\ \citenamefont {Qian}(2013)}]{Xu2013}%
  \BibitemOpen
  \bibfield  {author} {\bibinfo {author} {\bibfnamefont {X.}~\bibnamefont
  {Xu}}\ and\ \bibinfo {author} {\bibfnamefont {T.}~\bibnamefont {Qian}},\
  }\bibfield  {title} {\bibinfo {title} {Hydrodynamics of leidenfrost droplets
  in one-component fluids},\ }\href@noop {} {\bibfield  {journal} {\bibinfo
  {journal} {Phys. Rev. E}\ }\textbf {\bibinfo {volume} {87}},\ \bibinfo
  {pages} {1} (\bibinfo {year} {2013})}\BibitemShut {NoStop}%
\bibitem [{\citenamefont {Rueda~Villegas}\ \emph {et~al.}(2016)\citenamefont
  {Rueda~Villegas}, \citenamefont {Alis}, \citenamefont {Lepilliez},\ and\
  \citenamefont {Tanguy}}]{Rueda2016}%
  \BibitemOpen
  \bibfield  {author} {\bibinfo {author} {\bibfnamefont {L.}~\bibnamefont
  {Rueda~Villegas}}, \bibinfo {author} {\bibfnamefont {R.}~\bibnamefont
  {Alis}}, \bibinfo {author} {\bibfnamefont {M.}~\bibnamefont {Lepilliez}},\
  and\ \bibinfo {author} {\bibfnamefont {S.}~\bibnamefont {Tanguy}},\
  }\bibfield  {title} {\bibinfo {title} {A ghost fluid/level set method for
  boiling flows and liquid evaporation: Application to the leidenfrost
  effect},\ }\href@noop {} {\bibfield  {journal} {\bibinfo  {journal} {J.
  Comput. Phys.}\ }\textbf {\bibinfo {volume} {316}},\ \bibinfo {pages} {789}
  (\bibinfo {year} {2016})}\BibitemShut {NoStop}%
\bibitem [{\citenamefont {Rueda~Villegas}\ \emph {et~al.}(2017)\citenamefont
  {Rueda~Villegas}, \citenamefont {Tanguy}, \citenamefont {Castanet},
  \citenamefont {Caballina},\ and\ \citenamefont {Lemoine}}]{Rueda2017}%
  \BibitemOpen
  \bibfield  {author} {\bibinfo {author} {\bibfnamefont {L.}~\bibnamefont
  {Rueda~Villegas}}, \bibinfo {author} {\bibfnamefont {S.}~\bibnamefont
  {Tanguy}}, \bibinfo {author} {\bibfnamefont {G.}~\bibnamefont {Castanet}},
  \bibinfo {author} {\bibfnamefont {O.}~\bibnamefont {Caballina}},\ and\
  \bibinfo {author} {\bibfnamefont {F.}~\bibnamefont {Lemoine}},\ }\bibfield
  {title} {\bibinfo {title} {Direct numerical simulation of the impact of a
  droplet onto a hot surface above the leidenfrost temperature},\ }\href@noop
  {} {\bibfield  {journal} {\bibinfo  {journal} {Int. J. Heat Mass Transfer}\
  }\textbf {\bibinfo {volume} {104}},\ \bibinfo {pages} {1090} (\bibinfo {year}
  {2017})}\BibitemShut {NoStop}%
\bibitem [{\citenamefont {Guo}\ and\ \citenamefont {Cheng}(2019)}]{Guo2019}%
  \BibitemOpen
  \bibfield  {author} {\bibinfo {author} {\bibfnamefont {Q.}~\bibnamefont
  {Guo}}\ and\ \bibinfo {author} {\bibfnamefont {P.}~\bibnamefont {Cheng}},\
  }\bibfield  {title} {\bibinfo {title} {Direct numerical simulations of
  sessile droplet evaporation on a heated horizontal surface surrounded by
  moist air},\ }\href@noop {} {\bibfield  {journal} {\bibinfo  {journal} {Int.
  J. Heat Mass Transfer}\ }\textbf {\bibinfo {volume} {134}},\ \bibinfo {pages}
  {828} (\bibinfo {year} {2019})}\BibitemShut {NoStop}%
\bibitem [{\citenamefont {Chakraborty}\ \emph {et~al.}(2022)\citenamefont
  {Chakraborty}, \citenamefont {Chubynsky},\ and\ \citenamefont
  {Sprittles}}]{Chakraborty2022}%
  \BibitemOpen
  \bibfield  {author} {\bibinfo {author} {\bibfnamefont {I.}~\bibnamefont
  {Chakraborty}}, \bibinfo {author} {\bibfnamefont {M.}~\bibnamefont
  {Chubynsky}},\ and\ \bibinfo {author} {\bibfnamefont {J.}~\bibnamefont
  {Sprittles}},\ }\bibfield  {title} {\bibinfo {title} {Computational modelling
  of leidenfrost drops},\ }\href@noop {} {\bibfield  {journal} {\bibinfo
  {journal} {J. Fluid Mech.}\ }\textbf {\bibinfo {volume} {936}},\ \bibinfo
  {pages} {1} (\bibinfo {year} {2022})}\BibitemShut {NoStop}%
\bibitem [{\citenamefont {Mialhe}\ \emph {et~al.}(2023)\citenamefont {Mialhe},
  \citenamefont {Tanguy}, \citenamefont {Tranier}, \citenamefont {Popescu},\
  and\ \citenamefont {Legendre}}]{Mialhe2023}%
  \BibitemOpen
  \bibfield  {author} {\bibinfo {author} {\bibfnamefont {G.}~\bibnamefont
  {Mialhe}}, \bibinfo {author} {\bibfnamefont {S.}~\bibnamefont {Tanguy}},
  \bibinfo {author} {\bibfnamefont {L.}~\bibnamefont {Tranier}}, \bibinfo
  {author} {\bibfnamefont {E.-R.}\ \bibnamefont {Popescu}},\ and\ \bibinfo
  {author} {\bibfnamefont {D.}~\bibnamefont {Legendre}},\ }\bibfield  {title}
  {\bibinfo {title} {An extended model for the direct numerical simulation of
  droplet evaporation. influence of the marangoni convection on leidenfrost
  droplet},\ }\href@noop {} {\bibfield  {journal} {\bibinfo  {journal} {J.
  Comput. Phys.}\ }\textbf {\bibinfo {volume} {491}},\ \bibinfo {pages} {1}
  (\bibinfo {year} {2023})}\BibitemShut {NoStop}%
\bibitem [{\citenamefont {Biance}\ \emph {et~al.}(2003)\citenamefont {Biance},
  \citenamefont {Clanet},\ and\ \citenamefont {Quere}}]{Biance2003}%
  \BibitemOpen
  \bibfield  {author} {\bibinfo {author} {\bibfnamefont {A.-L.}\ \bibnamefont
  {Biance}}, \bibinfo {author} {\bibfnamefont {C.}~\bibnamefont {Clanet}},\
  and\ \bibinfo {author} {\bibfnamefont {D.}~\bibnamefont {Quere}},\ }\bibfield
   {title} {\bibinfo {title} {Leidenfrost drops},\ }\href@noop {} {\bibfield
  {journal} {\bibinfo  {journal} {Phys. Fluids}\ }\textbf {\bibinfo {volume}
  {15}},\ \bibinfo {pages} {1632} (\bibinfo {year} {2003})}\BibitemShut
  {NoStop}%
\bibitem [{\citenamefont {Myers}\ and\ \citenamefont
  {Charpin}(2009)}]{Myers2009}%
  \BibitemOpen
  \bibfield  {author} {\bibinfo {author} {\bibfnamefont {T.}~\bibnamefont
  {Myers}}\ and\ \bibinfo {author} {\bibfnamefont {J.}~\bibnamefont
  {Charpin}},\ }\bibfield  {title} {\bibinfo {title} {A mathematical model of
  the leidenfrost effect on an axisymmetric droplet},\ }\href@noop {}
  {\bibfield  {journal} {\bibinfo  {journal} {Phys. Fluids}\ }\textbf {\bibinfo
  {volume} {21}},\ \bibinfo {pages} {1} (\bibinfo {year} {2009})}\BibitemShut
  {NoStop}%
\bibitem [{\citenamefont {Shi}\ \emph {et~al.}(2019)\citenamefont {Shi},
  \citenamefont {Frank}, \citenamefont {Wang}, \citenamefont {Xu},
  \citenamefont {Lu},\ and\ \citenamefont {Grigoropoulos}}]{Shi2019}%
  \BibitemOpen
  \bibfield  {author} {\bibinfo {author} {\bibfnamefont {M.}~\bibnamefont
  {Shi}}, \bibinfo {author} {\bibfnamefont {F.}~\bibnamefont {Frank}}, \bibinfo
  {author} {\bibfnamefont {L.}~\bibnamefont {Wang}}, \bibinfo {author}
  {\bibfnamefont {F.}~\bibnamefont {Xu}}, \bibinfo {author} {\bibfnamefont
  {T.}~\bibnamefont {Lu}},\ and\ \bibinfo {author} {\bibfnamefont
  {C.}~\bibnamefont {Grigoropoulos}},\ }\bibfield  {title} {\bibinfo {title}
  {Role of jakob number in leidenfrost phenomena unveiled by theoretical
  modeling},\ }\href@noop {} {\bibfield  {journal} {\bibinfo  {journal} {Phys.
  Fluids}\ }\textbf {\bibinfo {volume} {31}},\ \bibinfo {pages} {1} (\bibinfo
  {year} {2019})}\BibitemShut {NoStop}%
\bibitem [{\citenamefont {Cai}\ \emph {et~al.}(2020)\citenamefont {Cai},
  \citenamefont {Mudawar}, \citenamefont {Liu},\ and\ \citenamefont
  {Si}}]{Cai2020}%
  \BibitemOpen
  \bibfield  {author} {\bibinfo {author} {\bibfnamefont {C.}~\bibnamefont
  {Cai}}, \bibinfo {author} {\bibfnamefont {I.}~\bibnamefont {Mudawar}},
  \bibinfo {author} {\bibfnamefont {H.}~\bibnamefont {Liu}},\ and\ \bibinfo
  {author} {\bibfnamefont {C.}~\bibnamefont {Si}},\ }\bibfield  {title}
  {\bibinfo {title} {Theoretical leidenfrost point (lfp) model for sessile
  droplet},\ }\href@noop {} {\bibfield  {journal} {\bibinfo  {journal} {J. Heat
  Mass Transfer}\ }\textbf {\bibinfo {volume} {146}},\ \bibinfo {pages} {1}
  (\bibinfo {year} {2020})}\BibitemShut {NoStop}%
\bibitem [{\citenamefont {Snoeijer}\ \emph {et~al.}(2009)\citenamefont
  {Snoeijer}, \citenamefont {Brunet},\ and\ \citenamefont
  {Eggers}}]{Snoeijer2009}%
  \BibitemOpen
  \bibfield  {author} {\bibinfo {author} {\bibfnamefont {J.}~\bibnamefont
  {Snoeijer}}, \bibinfo {author} {\bibfnamefont {P.}~\bibnamefont {Brunet}},\
  and\ \bibinfo {author} {\bibfnamefont {J.}~\bibnamefont {Eggers}},\
  }\bibfield  {title} {\bibinfo {title} {Maximum size of drops levitated by an
  air cushion},\ }\href@noop {} {\bibfield  {journal} {\bibinfo  {journal}
  {Phys. Rev. E}\ }\textbf {\bibinfo {volume} {79}},\ \bibinfo {pages} {1}
  (\bibinfo {year} {2009})}\BibitemShut {NoStop}%
\bibitem [{\citenamefont {Sobac}\ \emph {et~al.}(2014)\citenamefont {Sobac},
  \citenamefont {Rednikov}, \citenamefont {Dorbolo},\ and\ \citenamefont
  {Colinet}}]{Sobac2014}%
  \BibitemOpen
  \bibfield  {author} {\bibinfo {author} {\bibfnamefont {B.}~\bibnamefont
  {Sobac}}, \bibinfo {author} {\bibfnamefont {A.}~\bibnamefont {Rednikov}},
  \bibinfo {author} {\bibfnamefont {S.}~\bibnamefont {Dorbolo}},\ and\ \bibinfo
  {author} {\bibfnamefont {P.}~\bibnamefont {Colinet}},\ }\bibfield  {title}
  {\bibinfo {title} {Leidenfrost effect: Accurate drop shape modeling and
  refined scaling laws},\ }\href@noop {} {\bibfield  {journal} {\bibinfo
  {journal} {Phys. Rev. E}\ }\textbf {\bibinfo {volume} {90}},\ \bibinfo
  {pages} {1} (\bibinfo {year} {2014})}\BibitemShut {NoStop}%
\bibitem [{\citenamefont {Maquet}\ \emph {et~al.}(2015)\citenamefont {Maquet},
  \citenamefont {Brandenbourger}, \citenamefont {Sobac}, \citenamefont
  {Biance}, \citenamefont {Colinet},\ and\ \citenamefont
  {Dorbolo}}]{Maquet2015}%
  \BibitemOpen
  \bibfield  {author} {\bibinfo {author} {\bibfnamefont {L.}~\bibnamefont
  {Maquet}}, \bibinfo {author} {\bibfnamefont {M.}~\bibnamefont
  {Brandenbourger}}, \bibinfo {author} {\bibfnamefont {B.}~\bibnamefont
  {Sobac}}, \bibinfo {author} {\bibfnamefont {A.-L.}\ \bibnamefont {Biance}},
  \bibinfo {author} {\bibfnamefont {P.}~\bibnamefont {Colinet}},\ and\ \bibinfo
  {author} {\bibfnamefont {S.}~\bibnamefont {Dorbolo}},\ }\bibfield  {title}
  {\bibinfo {title} {Leidenfrost drops: Effect of gravity},\ }\href@noop {}
  {\bibfield  {journal} {\bibinfo  {journal} {Eur. Phys. Lett.}\ }\textbf
  {\bibinfo {volume} {110}},\ \bibinfo {pages} {1} (\bibinfo {year}
  {2015})}\BibitemShut {NoStop}%
\bibitem [{\citenamefont {Burton}\ \emph {et~al.}(2012)\citenamefont {Burton},
  \citenamefont {Sharpe}, \citenamefont {van~der Veen}, \citenamefont
  {Franco},\ and\ \citenamefont {Nagel}}]{Burton2012}%
  \BibitemOpen
  \bibfield  {author} {\bibinfo {author} {\bibfnamefont {J.}~\bibnamefont
  {Burton}}, \bibinfo {author} {\bibfnamefont {A.}~\bibnamefont {Sharpe}},
  \bibinfo {author} {\bibfnamefont {R.}~\bibnamefont {van~der Veen}}, \bibinfo
  {author} {\bibfnamefont {A.}~\bibnamefont {Franco}},\ and\ \bibinfo {author}
  {\bibfnamefont {S.}~\bibnamefont {Nagel}},\ }\bibfield  {title} {\bibinfo
  {title} {Geometry of the vapor layer under a leidenfrost drop},\ }\href@noop
  {} {\bibfield  {journal} {\bibinfo  {journal} {Phys. Rev. Lett.}\ }\textbf
  {\bibinfo {volume} {109}},\ \bibinfo {pages} {1} (\bibinfo {year}
  {2012})}\BibitemShut {NoStop}%
\bibitem [{\citenamefont {Zeuner}\ \emph {et~al.}(2019)\citenamefont {Zeuner},
  \citenamefont {Schwark}, \citenamefont {Hanisch},\ and\ \citenamefont
  {Ziese}}]{Ziese2019}%
  \BibitemOpen
  \bibfield  {author} {\bibinfo {author} {\bibfnamefont {M.}~\bibnamefont
  {Zeuner}}, \bibinfo {author} {\bibfnamefont {K.}~\bibnamefont {Schwark}},
  \bibinfo {author} {\bibfnamefont {C.}~\bibnamefont {Hanisch}},\ and\ \bibinfo
  {author} {\bibfnamefont {M.}~\bibnamefont {Ziese}},\ }\bibfield  {title}
  {\bibinfo {title} {Leidenfrost effect studied by video analysis},\
  }\href@noop {} {\bibfield  {journal} {\bibinfo  {journal} {Eur. J. Phys.}\
  }\textbf {\bibinfo {volume} {40}},\ \bibinfo {pages} {1} (\bibinfo {year}
  {2019})}\BibitemShut {NoStop}%
\bibitem [{\citenamefont {Orzechowski}(2019)}]{Orzechowski2021}%
  \BibitemOpen
  \bibfield  {author} {\bibinfo {author} {\bibfnamefont {T.}~\bibnamefont
  {Orzechowski}},\ }\bibfield  {title} {\bibinfo {title} {Peculiarities in
  leidenfrost water droplet evaporation},\ }\href@noop {} {\bibfield  {journal}
  {\bibinfo  {journal} {Heat Mass Transfer}\ }\textbf {\bibinfo {volume}
  {57}},\ \bibinfo {pages} {529} (\bibinfo {year} {2019})}\BibitemShut
  {NoStop}%
\bibitem [{\citenamefont {Sobac}\ \emph {et~al.}(2015)\citenamefont {Sobac},
  \citenamefont {Rednikov}, \citenamefont {Dorbolo},\ and\ \citenamefont
  {Colinet}}]{Sobac2015}%
  \BibitemOpen
  \bibfield  {author} {\bibinfo {author} {\bibfnamefont {B.}~\bibnamefont
  {Sobac}}, \bibinfo {author} {\bibfnamefont {A.}~\bibnamefont {Rednikov}},
  \bibinfo {author} {\bibfnamefont {S.}~\bibnamefont {Dorbolo}},\ and\ \bibinfo
  {author} {\bibfnamefont {P.}~\bibnamefont {Colinet}},\ }\href@noop {} {\emph
  {\bibinfo {title} {Droplet Wetting and Evaporation - Chapter 7: Leidenfrost
  Drops}}},\ edited by\ \bibinfo {editor} {\bibfnamefont {D.}~\bibnamefont
  {Brutin}}\ (\bibinfo  {publisher} {Academic Press},\ \bibinfo {address}
  {Oxford},\ \bibinfo {year} {2015})\ pp.\ \bibinfo {pages}
  {85--99}\BibitemShut {NoStop}%
\bibitem [{\citenamefont {Hashmi}\ \emph {et~al.}(2012)\citenamefont {Hashmi},
  \citenamefont {Xu}, \citenamefont {Coder}, \citenamefont {Osborne},
  \citenamefont {Spafford}, \citenamefont {Michael}, \citenamefont {Yu},\ and\
  \citenamefont {Xu}}]{Hashmi2012}%
  \BibitemOpen
  \bibfield  {author} {\bibinfo {author} {\bibfnamefont {A.}~\bibnamefont
  {Hashmi}}, \bibinfo {author} {\bibfnamefont {Y.}~\bibnamefont {Xu}}, \bibinfo
  {author} {\bibfnamefont {B.}~\bibnamefont {Coder}}, \bibinfo {author}
  {\bibfnamefont {P.}~\bibnamefont {Osborne}}, \bibinfo {author} {\bibfnamefont
  {J.}~\bibnamefont {Spafford}}, \bibinfo {author} {\bibfnamefont
  {G.}~\bibnamefont {Michael}}, \bibinfo {author} {\bibfnamefont
  {G.}~\bibnamefont {Yu}},\ and\ \bibinfo {author} {\bibfnamefont
  {J.}~\bibnamefont {Xu}},\ }\bibfield  {title} {\bibinfo {title} {Leidenfrost
  levitation: beyond droplets},\ }\href@noop {} {\bibfield  {journal} {\bibinfo
   {journal} {Sci. Rep.}\ }\textbf {\bibinfo {volume} {2}},\ \bibinfo {pages}
  {1} (\bibinfo {year} {2012})}\BibitemShut {NoStop}%
\bibitem [{\citenamefont {Sugioka}\ and\ \citenamefont
  {Segawa}(2018)}]{Segawa2018}%
  \BibitemOpen
  \bibfield  {author} {\bibinfo {author} {\bibfnamefont {H.}~\bibnamefont
  {Sugioka}}\ and\ \bibinfo {author} {\bibfnamefont {S.}~\bibnamefont
  {Segawa}},\ }\bibfield  {title} {\bibinfo {title} {Controllable leidenfrost
  glider on a shallow water layer},\ }\href@noop {} {\bibfield  {journal}
  {\bibinfo  {journal} {AIP Adv.}\ }\textbf {\bibinfo {volume} {8}},\ \bibinfo
  {pages} {1} (\bibinfo {year} {2018})}\BibitemShut {NoStop}%
\bibitem [{\citenamefont {Ok}\ \emph {et~al.}(2011)\citenamefont {Ok},
  \citenamefont {Lopez-Oña}, \citenamefont {Nikitopoulos}, \citenamefont
  {Wong},\ and\ \citenamefont {Park}}]{Park2011}%
  \BibitemOpen
  \bibfield  {author} {\bibinfo {author} {\bibfnamefont {J.}~\bibnamefont
  {Ok}}, \bibinfo {author} {\bibfnamefont {E.}~\bibnamefont {Lopez-Oña}},
  \bibinfo {author} {\bibfnamefont {D.}~\bibnamefont {Nikitopoulos}}, \bibinfo
  {author} {\bibfnamefont {H.}~\bibnamefont {Wong}},\ and\ \bibinfo {author}
  {\bibfnamefont {S.}~\bibnamefont {Park}},\ }\bibfield  {title} {\bibinfo
  {title} {Propulsion of droplets on micro- and sub-micron ratchet surfaces in
  the leidenfrost temperature regime},\ }\href@noop {} {\bibfield  {journal}
  {\bibinfo  {journal} {Microfluid. Nanofluid.}\ }\textbf {\bibinfo {volume}
  {10}},\ \bibinfo {pages} {1045} (\bibinfo {year} {2011})}\BibitemShut
  {NoStop}%
\bibitem [{\citenamefont {Adera}\ \emph {et~al.}(2013)\citenamefont {Adera},
  \citenamefont {Raj}, \citenamefont {Enright},\ and\ \citenamefont
  {Wang}}]{Adera2013}%
  \BibitemOpen
  \bibfield  {author} {\bibinfo {author} {\bibfnamefont {S.}~\bibnamefont
  {Adera}}, \bibinfo {author} {\bibfnamefont {R.}~\bibnamefont {Raj}}, \bibinfo
  {author} {\bibfnamefont {R.}~\bibnamefont {Enright}},\ and\ \bibinfo {author}
  {\bibfnamefont {E.}~\bibnamefont {Wang}},\ }\bibfield  {title} {\bibinfo
  {title} {Non-wetting droplets on hot superhydrophilic surfaces},\ }\href@noop
  {} {\bibfield  {journal} {\bibinfo  {journal} {Nat. Commun.}\ }\textbf
  {\bibinfo {volume} {4}},\ \bibinfo {pages} {1} (\bibinfo {year}
  {2013})}\BibitemShut {NoStop}%
\bibitem [{\citenamefont {Dodd}\ \emph {et~al.}(2016)\citenamefont {Dodd},
  \citenamefont {Wood}, \citenamefont {Geraldi}, \citenamefont {Wells},
  \citenamefont {McHale}, \citenamefont {Xu}, \citenamefont {Stuart-Cole},
  \citenamefont {Martin},\ and\ \citenamefont {Newton}}]{Dodd2016}%
  \BibitemOpen
  \bibfield  {author} {\bibinfo {author} {\bibfnamefont {L.}~\bibnamefont
  {Dodd}}, \bibinfo {author} {\bibfnamefont {D.}~\bibnamefont {Wood}}, \bibinfo
  {author} {\bibfnamefont {N.}~\bibnamefont {Geraldi}}, \bibinfo {author}
  {\bibfnamefont {G.}~\bibnamefont {Wells}}, \bibinfo {author} {\bibfnamefont
  {G.}~\bibnamefont {McHale}}, \bibinfo {author} {\bibfnamefont
  {B.}~\bibnamefont {Xu}}, \bibinfo {author} {\bibfnamefont {S.}~\bibnamefont
  {Stuart-Cole}}, \bibinfo {author} {\bibfnamefont {J.}~\bibnamefont
  {Martin}},\ and\ \bibinfo {author} {\bibfnamefont {M.}~\bibnamefont
  {Newton}},\ }\bibfield  {title} {\bibinfo {title} {Low friction droplet
  transportation on a substrate with a selective leidenfrost effect},\
  }\href@noop {} {\bibfield  {journal} {\bibinfo  {journal} {ACS Appl. Mater.
  Interfaces}\ }\textbf {\bibinfo {volume} {8}},\ \bibinfo {pages} {22658}
  (\bibinfo {year} {2016})}\BibitemShut {NoStop}%
\bibitem [{\citenamefont {Dodd}\ \emph {et~al.}(2020)\citenamefont {Dodd},
  \citenamefont {Agrawal}, \citenamefont {Geraldi}, \citenamefont {Xu},
  \citenamefont {Wells}, \citenamefont {Martin}, \citenamefont {Newton},
  \citenamefont {McHale},\ and\ \citenamefont {Wood}}]{Dodd2020}%
  \BibitemOpen
  \bibfield  {author} {\bibinfo {author} {\bibfnamefont {L.}~\bibnamefont
  {Dodd}}, \bibinfo {author} {\bibfnamefont {P.}~\bibnamefont {Agrawal}},
  \bibinfo {author} {\bibfnamefont {N.}~\bibnamefont {Geraldi}}, \bibinfo
  {author} {\bibfnamefont {B.}~\bibnamefont {Xu}}, \bibinfo {author}
  {\bibfnamefont {G.}~\bibnamefont {Wells}}, \bibinfo {author} {\bibfnamefont
  {J.}~\bibnamefont {Martin}}, \bibinfo {author} {\bibfnamefont
  {M.}~\bibnamefont {Newton}}, \bibinfo {author} {\bibfnamefont
  {G.}~\bibnamefont {McHale}},\ and\ \bibinfo {author} {\bibfnamefont
  {D.}~\bibnamefont {Wood}},\ }\bibfield  {title} {\bibinfo {title} {Planar
  selective leidenfrost propulsion without physically structured substrates or
  walls},\ }\href@noop {} {\bibfield  {journal} {\bibinfo  {journal} {Appl.
  Phys. Lett.}\ }\textbf {\bibinfo {volume} {117}},\ \bibinfo {pages} {1}
  (\bibinfo {year} {2020})}\BibitemShut {NoStop}%
\bibitem [{\citenamefont {Liang}\ and\ \citenamefont
  {Mudawar}(2017{\natexlab{a}})}]{Mudawar2017a}%
  \BibitemOpen
  \bibfield  {author} {\bibinfo {author} {\bibfnamefont {G.}~\bibnamefont
  {Liang}}\ and\ \bibinfo {author} {\bibfnamefont {I.}~\bibnamefont
  {Mudawar}},\ }\bibfield  {title} {\bibinfo {title} {Review of spray cooling
  – part 1: Single-phase and nucleate boiling regimes, and critical heat
  flux},\ }\href@noop {} {\bibfield  {journal} {\bibinfo  {journal} {Int. J.
  Heat Mass Transfer}\ }\textbf {\bibinfo {volume} {115}},\ \bibinfo {pages}
  {1174} (\bibinfo {year} {2017}{\natexlab{a}})}\BibitemShut {NoStop}%
\bibitem [{\citenamefont {Liang}\ and\ \citenamefont
  {Mudawar}(2017{\natexlab{b}})}]{Mudawar2017b}%
  \BibitemOpen
  \bibfield  {author} {\bibinfo {author} {\bibfnamefont {G.}~\bibnamefont
  {Liang}}\ and\ \bibinfo {author} {\bibfnamefont {I.}~\bibnamefont
  {Mudawar}},\ }\bibfield  {title} {\bibinfo {title} {Review of spray cooling
  – part 2: High temperature boiling regimes and quenching applications},\
  }\href@noop {} {\bibfield  {journal} {\bibinfo  {journal} {Int. J. Heat Mass
  Transfer}\ }\textbf {\bibinfo {volume} {115}},\ \bibinfo {pages} {1206}
  (\bibinfo {year} {2017}{\natexlab{b}})}\BibitemShut {NoStop}%
\bibitem [{\citenamefont {de~Gennes}\ \emph {et~al.}(2003)\citenamefont
  {de~Gennes}, \citenamefont {Brochard-Wyart},\ and\ \citenamefont
  {Quere}}]{deGennes}%
  \BibitemOpen
  \bibfield  {author} {\bibinfo {author} {\bibfnamefont {P.-G.}\ \bibnamefont
  {de~Gennes}}, \bibinfo {author} {\bibfnamefont {F.}~\bibnamefont
  {Brochard-Wyart}},\ and\ \bibinfo {author} {\bibfnamefont {D.}~\bibnamefont
  {Quere}},\ }\href@noop {} {\emph {\bibinfo {title} {Capillarity and wetting
  phenomena: Drops, Bubbles, Pearls, Waves}}}\ (\bibinfo  {publisher}
  {Springer},\ \bibinfo {year} {2003})\BibitemShut {NoStop}%
\bibitem [{Sup()}]{SuppMat}%
  \BibitemOpen
  \href@noop {} {\bibinfo {title} {See supplementary material at [url will be
  inserted by publisher] for further details on the procedures involving the
  hankel transform, applied to find the dimensionless velocity, pressure and
  temperature fields, the proposed dimensionless functions for the boundary
  conditions, the determination of the mass fraction of vapor at the air-liquid
  interface, and some examples of the thermophysical properties, characteristic
  parameters and dimensionless numbers of the fluids}},\ \bibinfo
  {howpublished} {\url{https://www.}},\ \bibinfo {note} {includes Refs.
  [25--26,51,54--55].}\BibitemShut {Stop}%
\bibitem [{\citenamefont {Sneddon}(1995)}]{Sneddon}%
  \BibitemOpen
  \bibfield  {author} {\bibinfo {author} {\bibfnamefont {I.}~\bibnamefont
  {Sneddon}},\ }\href@noop {} {\emph {\bibinfo {title} {Fourier transforms}}}\
  (\bibinfo  {publisher} {Dover},\ \bibinfo {year} {1995})\BibitemShut
  {NoStop}%
\bibitem [{\citenamefont {Zaitsev}\ \emph {et~al.}(2017)\citenamefont
  {Zaitsev}, \citenamefont {Kirichenko}, \citenamefont {Ajaev},\ and\
  \citenamefont {Kabov}}]{Kabov2017}%
  \BibitemOpen
  \bibfield  {author} {\bibinfo {author} {\bibfnamefont {D.}~\bibnamefont
  {Zaitsev}}, \bibinfo {author} {\bibfnamefont {D.}~\bibnamefont {Kirichenko}},
  \bibinfo {author} {\bibfnamefont {V.}~\bibnamefont {Ajaev}},\ and\ \bibinfo
  {author} {\bibfnamefont {O.}~\bibnamefont {Kabov}},\ }\bibfield  {title}
  {\bibinfo {title} {Levitation and self-organization of liquid microdroplets
  over dry heated substrates},\ }\href@noop {} {\bibfield  {journal} {\bibinfo
  {journal} {Phys. Rev. Lett.}\ }\textbf {\bibinfo {volume} {119}},\ \bibinfo
  {pages} {1} (\bibinfo {year} {2017})}\BibitemShut {NoStop}%
\bibitem [{\citenamefont {Alassar}(1999)}]{Alassar1999}%
  \BibitemOpen
  \bibfield  {author} {\bibinfo {author} {\bibfnamefont {R.}~\bibnamefont
  {Alassar}},\ }\bibfield  {title} {\bibinfo {title} {Heat conduction from
  spheroids},\ }\href@noop {} {\bibfield  {journal} {\bibinfo  {journal} {J.
  Heat Transfer}\ }\textbf {\bibinfo {volume} {121}},\ \bibinfo {pages} {497}
  (\bibinfo {year} {1999})}\BibitemShut {NoStop}%
\bibitem [{\citenamefont {Poling}\ \emph {et~al.}(2001)\citenamefont {Poling},
  \citenamefont {Prausnitz},\ and\ \citenamefont {O'Conell}}]{Poling}%
  \BibitemOpen
  \bibfield  {author} {\bibinfo {author} {\bibfnamefont {B.}~\bibnamefont
  {Poling}}, \bibinfo {author} {\bibfnamefont {J.}~\bibnamefont {Prausnitz}},\
  and\ \bibinfo {author} {\bibfnamefont {J.}~\bibnamefont {O'Conell}},\
  }\href@noop {} {\emph {\bibinfo {title} {The properties of gases and
  liquids}}},\ \bibinfo {edition} {5th}\ ed.\ (\bibinfo  {publisher}
  {McGraw-Hill},\ \bibinfo {year} {2001})\BibitemShut {NoStop}%
\bibitem [{\citenamefont {Yaws}(1999)}]{Yaws}%
  \BibitemOpen
  \bibfield  {author} {\bibinfo {author} {\bibfnamefont {C.}~\bibnamefont
  {Yaws}},\ }\href@noop {} {\emph {\bibinfo {title} {Chemical properties
  handbook: Physical, thermodynamic, environmental, transport, safety, and
  health related properties for organic and inorganic chemicals}}}\ (\bibinfo
  {publisher} {McGraw-Hill},\ \bibinfo {year} {1999})\BibitemShut {NoStop}%
\bibitem [{\citenamefont {Yang}\ \emph {et~al.}(2023)\citenamefont {Yang},
  \citenamefont {Li}, \citenamefont {Wang}, \citenamefont {Fan}, \citenamefont
  {Ma}, \citenamefont {Yu}, \citenamefont {Guo}, \citenamefont {Chen},
  \citenamefont {Wang},\ and\ \citenamefont {Deng}}]{Yang2023}%
  \BibitemOpen
  \bibfield  {author} {\bibinfo {author} {\bibfnamefont {J.}~\bibnamefont
  {Yang}}, \bibinfo {author} {\bibfnamefont {Y.}~\bibnamefont {Li}}, \bibinfo
  {author} {\bibfnamefont {D.}~\bibnamefont {Wang}}, \bibinfo {author}
  {\bibfnamefont {Y.}~\bibnamefont {Fan}}, \bibinfo {author} {\bibfnamefont
  {Y.}~\bibnamefont {Ma}}, \bibinfo {author} {\bibfnamefont {F.}~\bibnamefont
  {Yu}}, \bibinfo {author} {\bibfnamefont {J.}~\bibnamefont {Guo}}, \bibinfo
  {author} {\bibfnamefont {L.}~\bibnamefont {Chen}}, \bibinfo {author}
  {\bibfnamefont {Z.}~\bibnamefont {Wang}},\ and\ \bibinfo {author}
  {\bibfnamefont {X.}~\bibnamefont {Deng}},\ }\bibfield  {title} {\bibinfo
  {title} {A standing leidenfrost drop with sufi whirling},\ }\href@noop {}
  {\bibfield  {journal} {\bibinfo  {journal} {Proc. Natl. Acad. Sci.}\ }\textbf
  {\bibinfo {volume} {120}},\ \bibinfo {pages} {1} (\bibinfo {year}
  {2023})}\BibitemShut {NoStop}%
\end{thebibliography}%

\end{document}